\documentclass[11pt,a4paper]{emulateapj}
\newcommand{\p}{\partial}
\newcommand{\M}{{\cal M}}

\usepackage{epsfig}
\usepackage{amsmath}
\usepackage{natbib}
\usepackage{verbatim}

\slugcomment{Accepted for publication in ApJ}
\shorttitle{MHD SASI}
\shortauthors{Guilet \& Foglizzo}

\begin{document}

\title{Toward a magnetohydrodynamic theory of the stationary accretion shock instability: \\  toy model of the advective-acoustic cycle in a magnetized flow}

\author{J\'er\^ome Guilet  and Thierry Foglizzo}
\affil{Laboratoire AIM, CEA/DSM-CNRS-Universit\'e Paris Diderot, IRFU/Service d'Astrophysique, \\
CEA-Saclay F-91191 Gif-sur-Yvette, France. }
\email{jerome.guilet@cea.fr}

\begin{abstract}

The effect  of a magnetic field on the linear phase of the advective-acoustic instability is investigated, as a first step toward a magnetohydrodynamic (MHD) theory of the stationary accretion shock instability taking place during stellar core collapse. We study a toy model where the flow behind a planar stationary accretion shock is adiabatically decelerated by an external potential. Two magnetic field geometries are considered: parallel or perpendicular to the shock. The entropyÐvorticity wave, which is simply advected in the unmagnetized limit, separates into five different waves: the entropy perturbations are advected, while the vorticity can propagate along the field lines through two Alfv\'en waves and two slow magnetosonic waves. The two cycles existing in the unmagnetized limit, advectiveÐacoustic and purely acoustic, are replaced by up to six distinct MHD cycles. The phase differences among the cycles play an important role in determining the total cycle efficiency and hence the growth rate. Oscillations in the growth rate as a function of the magnetic field strength are due to this varying phase shift. A vertical magnetic field hardly affects the cycle efficiency in the regime of super-Alfv\'enic accretion that is considered. In contrast, we find that a horizontal magnetic field strongly increases the efficiencies of the vorticity cycles that bend the field lines, resulting in a significant increase of the growth rate if the different cycles are in phase. These magnetic effects are significant for large-scale modes if the Alfv\'en velocity is a sizable fraction of the flow velocity.

\end{abstract}

\keywords{instabilities --- shock waves --- magnetic fields---magnetohydrodynamics---supernovae: general}

\section{Introduction}

The delayed energy deposition by neutrinos may be a viable explosion mechanism of massive stars \citep{bethe85} if multidimensional hydrodynamical instabilities are taken into account during the first second after shock bounce, in the inner 200km of the massive star \citep{buras06b,marek09}. By capturing some gas in large scale convective cells, these instabilities increase the exposure time of this gas to the heating by neutrinos \citep{marek09,murphy08b,fernandez09b}. The $l=1,2$ deformation of the shock seems to be mainly due to the Standing Accretion Shock Instability (SASI), discovered by \cite{blondin03}. 
This asymmetric instability may influence the kick and spin of the residual pulsar \citep{scheck06,blondin07a}, trigger the excitation of g-modes \citep{burrows06} and the emission of gravitational waves \citep{marek09b,ott09}. These potential consequences of SASI are based on numerical simulations where the magnetic field is neglected, because the magnetic pressure is thought to be small compared to the thermal pressure in a typical core collapse. Other models of core collapse supernovae, in which magnetic forces play a dynamical role, usually rely on a fast rotation rate from which the magnetic energy is extracted by field winding and local instabilities \citep{akiyama03,moiseenko06,shibata06,burrows07b,obergaulinger09}. Surprisingly, \cite{endeve08} argued that the turbulent flow associated with SASI may be able to amplify an initially weak magnetic field to a dynamically significant strength of $ 10^{15} $G, even if the progenitor is not rotating. This latter result, if confirmed, shows the importance of understanding SASI in the context of MHD even if the progenitor is weakly magnetized. 

As a first step in this direction, we investigate the linear phase of SASI in a moderate magnetic field.
The linear mechanism of SASI has been identified as an advective-acoustic cycle by \cite{blondin03,ohnishi06,foglizzo07,scheck08,fernandez09a}. The interplay of acoustic waves and advected perturbations of entropy/vorticity has been illustrated using a simple toy model by \cite{foglizzo09} and \cite{sato09}, in which the flow is planar and adiabatic. A similar cycle is expected to exist in a MHD flow, but should be modified by the magnetic field depending on its strength and direction, and could potentially involve entropy waves, fast and slow magnetosonic waves, and Alfv\'en waves. 
The local stability of an isolated fast MHD shock has been established by \cite{gardner64}, and its interaction with MHD waves has been characterized by \cite{mckenzie70}. However, the global stability of the cycles resulting from the linear coupling of these MHD waves through a region of deceleration below the shock has never been studied. For this purpose we extend the toy model of \cite{foglizzo09} to a magnetized fluid, and investigate two possible orientations of the magnetic field.

It should be noted that the existence of SASI type instabilities in a magnetized flow has been recently proposed by \cite{camus09} to explain the oscillations of the termination shock in their MHD simulations of the relativistic pulsar wind in the Crab nebula.
The present study can thus be considered more generally as a first step toward understanding the MHD extension of the classical advective-acoustic cycle.

In Sect.~2 we describe the stationary flow of our MHD toy model and define the method of our linear analysis. The efficiencies of all existing cycles are calculated in Sect.~3, and the growth rate of the resulting eigenmode is computed in Sect.~4. The results are summarized and discussed in Sect.~5. Most algebraic derivations have been separated from the main text for the sake of clarity, and can be found in the Appendices.

\section{Set up: the magnetic toy model}

\subsection{The stationary flow}

\begin{figure}[tbp]
\begin{center}
\includegraphics[width=\columnwidth]{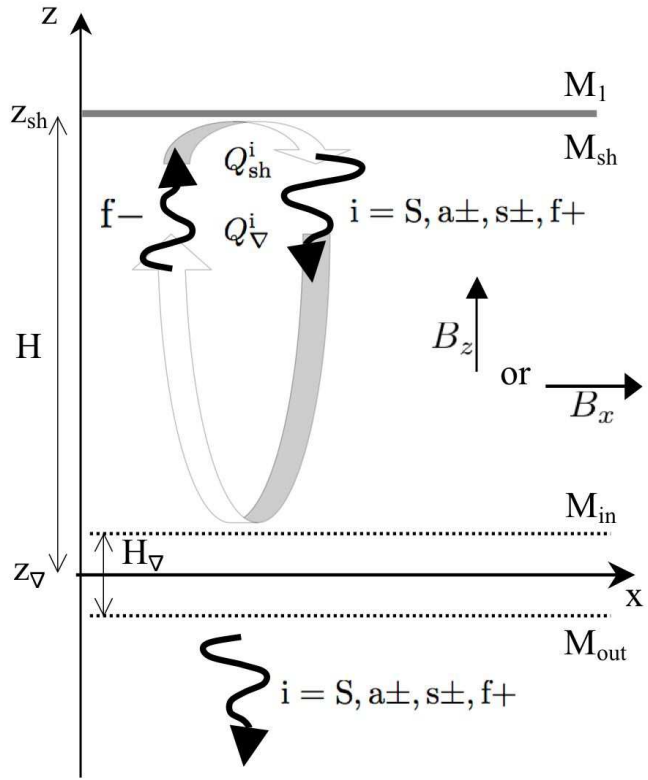}
\caption{Schematic view of the toy model. A magnetic field that is either vertical or horizontal is added to the hydrodynamical toy model of \cite{foglizzo09}. Six waves propagate from the shock to the inhomogeneous region near $z_{\nabla}$ (an entropy wave ${\rm S}$, two Alfv\'en waves ${\rm a}\pm$, two slow waves ${\rm s}\pm$ and a fast wave ${\rm f}+$) and couple to the upward propagating fast wave ${\rm f}-$. The linear coupling between a wave $i$ and the wave ${\rm f}-$ is described by the coupling coefficients $Q^{i}_{\nabla}$ and $Q^{i}_{\rm sh}$ defined in Sec.~2.2}
\label{toymodel}
\end{center}
\end{figure}

In the toy model of the advective-acoustic instability introduced by \cite{foglizzo09}, an ideal gas characterized by an adiabatic index $\gamma=4/3$ flows in the negative $z$-direction. This planar adiabatic flow is decelerated through a stationary shock at $z=z_{\rm sh} $, with an incident Mach number $\M_1$. It is further decelerated over a distance $H_\nabla<H$ in the vicinity of $z_\nabla\equiv z_{\rm sh}-H$, by a decrease $\Delta \Phi $ of the external potential :
\begin{equation}
\Phi\left(z\right) \equiv \frac{\Delta \Phi}{2}\left\lbrack \tanh\left( \frac{z-z_{\nabla}}{H_{\nabla}/2}\right) + 1 \right\rbrack. 
\end{equation}
We add to this flow a magnetic field $(B_x,0,B_z)$ that is oriented either vertically ($B_x=0$), or horizontally ($B_z=0$), as illustrated in Fig.~\ref{toymodel}. This magnetic field does not depend on $x$ or $y$.

Let us denote by $P$ the gas pressure, $\rho$ the density, $c\equiv \sqrt{\gamma P/\rho} $ the sound speed, and $v$ the fluid velocity. $S \equiv  \left\lbrack \log\left( \left(P/P_{\rm sh}\right)/ \left(\rho/ \rho_{\rm sh}\right)^{ \gamma}\right)  \right\rbrack/\left( \gamma -1\right) $ is a dimensionless measure of the entropy. $v_{{\rm A}i} \equiv \sqrt{ B_{i}^{2}/ \mu_{0} \rho }$ is the Alfv\'en speed along the direction $i=x$ or $z$ .

The stationary flow is governed by the conservation of the mass flux $\rho v_{z}$, the energy density $v_{z}^{2}/2 + c^{2}/\left(\gamma - 1\right)  + v_{{\rm A}x}^{2} + \Phi$, and the entropy $S$. The conservation of $B_{z}$  and $B_{x}/\rho$ follows respectively from the conservation of the magnetic flux and the frozen-in field lines in ideal MHD. 

A generalization of the Rankine-Hugoniot jump conditions to MHD characterizes the post-shock flow as a function of the upstream conditions (Eqs.~(\ref{machB}-\ref{chi}) in Appendix~A). The additional pressure due to a horizontal magnetic field decreases the compression at the shock and increases the post shock Mach number  by less than $10\%$ if  $v_{\rm Ash}<v_{\rm sh}$.

\begin{figure}[tbp]
\begin{center}
\includegraphics[width=\columnwidth]{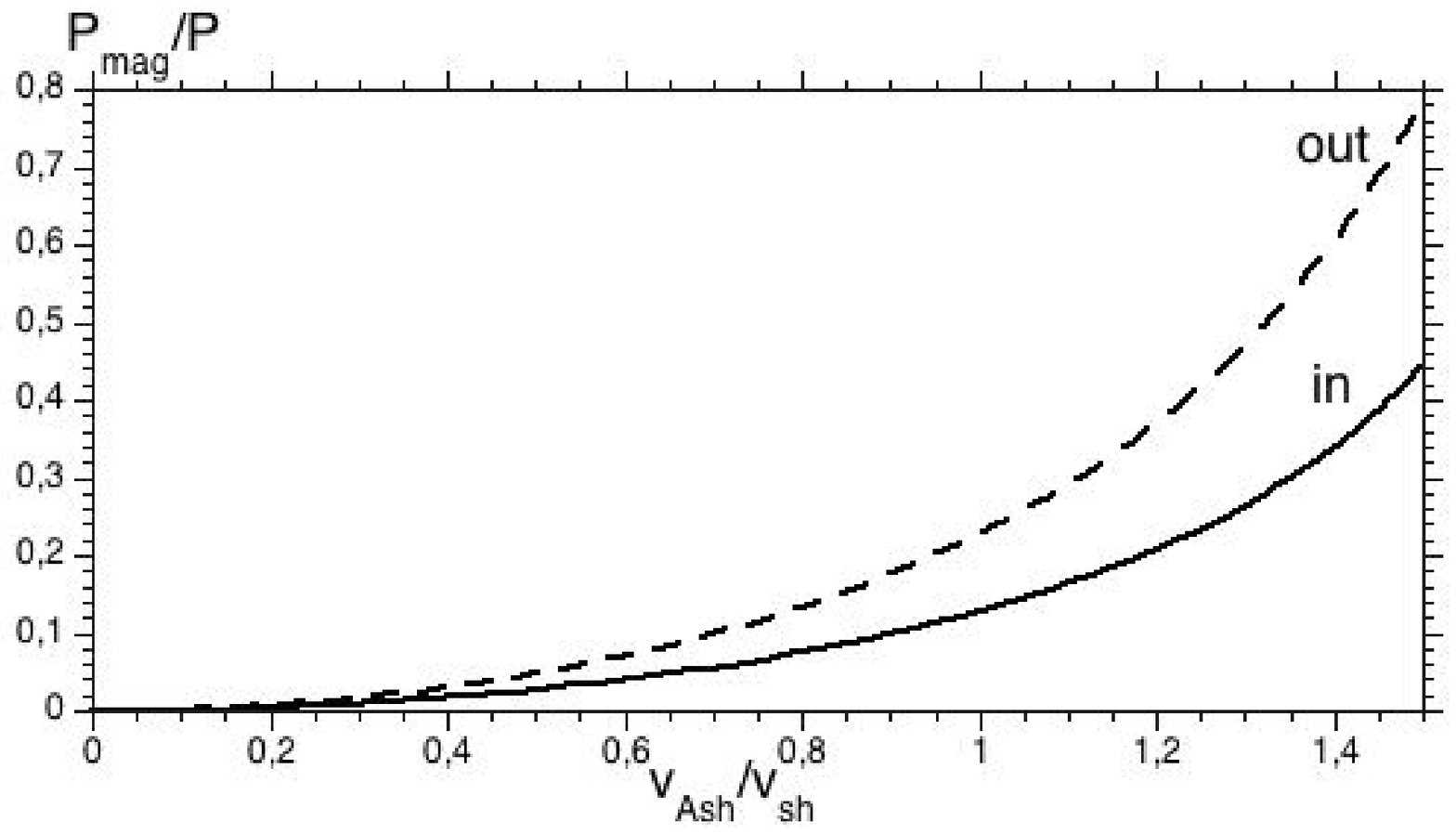}
\includegraphics[width=\columnwidth]{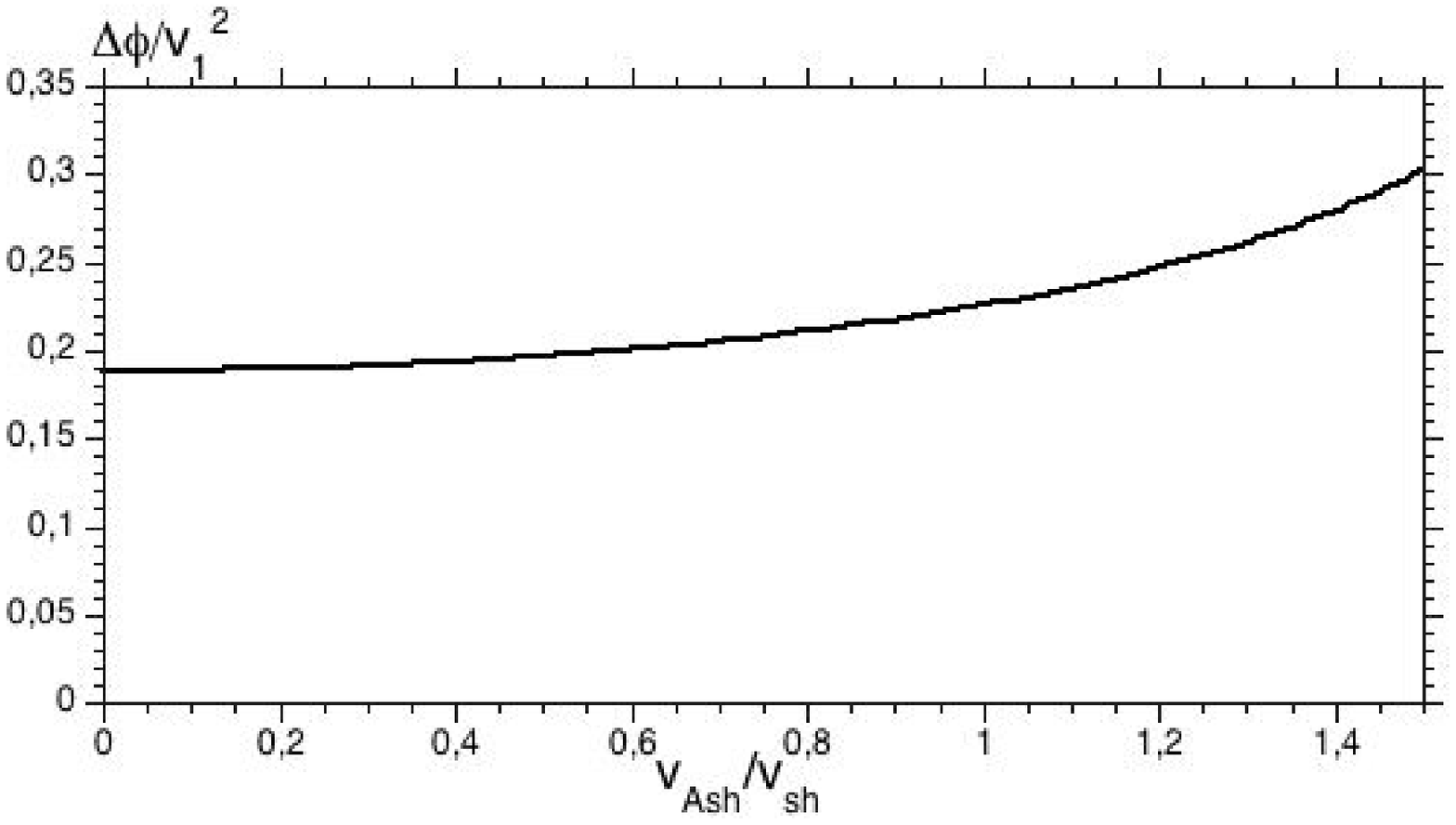}
\caption{Effect of the magnetic field on the stationary flow. \emph{Upper panel:} Ratio of magnetic pressure to thermal pressure just below the shock (full line) and below the potential jump (dashed line). \emph{Bottom panel:} Change of the potential jump if the compression is kept constant with $c_{\rm in}^{2}/c_{\rm out}^{2} = 0.75$. The potential jump is increased by $\sim 20\%$ at $v_{\rm A}=1$. }
\label{cinout}
\end{center}
\end{figure}

The flow below the adiabatic region of deceleration is characterized by the size $\Delta \Phi$ of the potential jump, or equivalently by the ratio of sound speed $c_{\rm in}/c_{\rm out}$, using the conservation of mass, energy and magnetic fluxes:
\begin{eqnarray}
\frac{B_{x\rm in}}{B_{x\rm out}}  &=&   \left( \frac{c_{\rm in}}{c_{\rm out}} \right)^{2\over \gamma -1}  ,  \\
\frac{v_{\rm in}}{v_{\rm out}} & = &   \left( \frac{c_{\rm in}}{c_{\rm out}} \right)^{-{2\over\gamma -1}}   ,  \\
\frac{\M_{\rm in}}{\M_{\rm out}} & = &    \left( \frac{c_{\rm in}}{c_{\rm out}} \right)^{-{\gamma + 1\over\gamma -1}}  ,  \\
\Delta \Phi &=&\left\lbrack \frac{v^{2}}{2} + v_{\rm A}^{2} + \frac{c^{2}}{\gamma - 1} \right\rbrack^{\rm out}_{\rm in}.
\end{eqnarray}
Unless otherwise specified, we use the values of the parameters $c_{\rm in}^2/c_{\rm out}^2 = 0.75$, $H_{\nabla}/H = 10^{-3}$, $\M_{1} = 5$ when studying the effect of the magnetic field. This choice implies an increase of the height of the potential jump with the field strength, which becomes significant for $v_{{\rm A}x{\rm sh}}/v_{\rm sh} > 1-1.5 $, i.e. when the magnetic pressure becomes comparable with the thermal pressure (Fig.~\ref{cinout}). Alternatively, one could choose to keep the potential jump constant. This choice of parameterization is unimportant in the present study, which is focussed on the regime of weak magnetic pressure.

\subsection{Linear perturbations}

The 1-D stationary flow is perturbed in 3-D, using Fourier transforms in time and in the directions $x$ and $y$ where the boundary conditions are periodic. The structure of the perturbation of a physical quantity $f$ is thus: $ \delta f(z)\times\exp i\left(k_{x}x + k_{y}y - \omega t  \right) $. For a given horizontal size of the computational domain $L_x$ and $L_y$ (we chose $L_{x}=L_{y}=4H$), the horizontal wavenumbers take discrete values $k_{i} \equiv 2\pi n_{i}/L_{i}$, where $i=x,y$ and $n_{i}$ is the number of horizontal wavelengths in the $i$-direction. We impose a leaking boundary condition below the potential jump (i.e. no wave propagating upward from below), and leave the supersonic flow unperturbed ahead of the shock. The corresponding equations are detailed in Appendices A and B.

In each uniform section of the flow, perturbations can be decomposed into 7 waves: two fast magnetosonic, two slow magnetosonic, two Alfv\'en and one entropy wave. If the shock is a fast MHD shock, only the fast magnetosonic wave is able to reach it from below. We do not consider a slow MHD shock because it would require Alfv\'en waves to be able to propagate upstream of the shock (which would correspond to an unrealistically strong magnetic field), nor transAlfv\'enic discontinuities, which have been shown to be non evolutionary  \citep{syrovatskii59}. As a consequence, we always consider a situation where, due to advection, 6 of the 7 waves are propagating downward in the post-shock region (Fig.~\ref{toymodel}). 

A global mode can be decomposed into 6 cycles, associated with each of the six waves propagating downward, while the upward propagation always corresponds to a fast magnetosonic wave. Waves are linearly coupled at the shock and in the region of flow gradients around $z_\nabla$. Following \cite{foglizzo09}, the coupling efficiency $Q_{\nabla}^{i}$ describes the ratio of amplitudes of a perturbation $\delta A$ of a physical quantity $A$ which is propagated from $z_{\rm sh}$ to $z_{\nabla}$, generates a fast magnetosonic wave in the region of gradients, which propagates back to the shock:
\begin{eqnarray}
Q_{\nabla}^{i} &\equiv &{ \delta A^{{\rm f}-}_{\rm sh}\over \delta A^{i}_{\rm sh}},\\
 & &i={\rm s}\pm,{\rm f}+,{\rm a}\pm,{\rm S},\nonumber
\end{eqnarray}
where the symbol $+$ ($-$) is used for waves propagating downward (upward) in the frame advected with the fluid. The superscript ${\rm s}\pm$ stands for slow magnetosonic waves, ${\rm f}\pm$ for the fast ones, ${\rm a}\pm$ for the Alfv\'en waves and ${\rm S}$ for the entropy wave. 

The coupling efficiency at the shock $Q_{\rm sh}^{i}$ measures the creation of a wave $i$ when the fast magnetosonic wave ${\rm f}-$ perturbs the shock:
\begin{equation}
Q_{\rm sh}^{i} \equiv  {\delta A^{i}_{\rm sh}\over \delta A^{{\rm f}-}_{\rm sh}}.
\end{equation}
The choice of the physical quantity $A$ is driven by the simplicity of the resulting equations (see Appendices A and B). The total cycle efficiency $Q^{i}$, defined as follows, does not depend on this choice:
\begin{equation}
Q^{i} \equiv Q_{\nabla}^{i}Q_{\rm sh}^{i}.
\end{equation}

We define $Q_{\rm tot}$ as the sum of all the cycles constants, and $Q_{\rm vort} \equiv Q^{{\rm s}+} + Q^{{\rm s}-} + Q^{{\rm a}+} + Q^{{\rm a}-}  $ is restricted to the cycles involving vorticity. 

It is instructive to study how these six cycles relate to the two hydrodynamical cycles, by considering the weak field limit. Since the acoustic cycle straightforwardly becomes the fast magnetosonic cycle, its reflection efficiency is noted $R$ instead of $Q^{{\rm f}+}$. The advective-acoustic cycle splits between an entropy-fast cycle where the entropy wave is still advected, and four Alfv\'en-fast, slow-fast cycles that contain the vorticity, which is no longer passively advected. In the setup we consider, the entropy cycle typically has an efficiency of $Q_{\rm S} \sim 1$ while the vorticity cycle has an efficiency of $Q_{\rm vort} \sim 0.3$ in the hydrodynamical limit. How is vorticity distributed between the Alfv\'en and slow waves ? The distinction between slow and Alfv\'en waves at the shock can be anticipated by remembering that the velocity and magnetic field perturbations of Alfv\'en waves are along the direction $\bf{k}\times\bf{B}$, while that of slow waves is in the plane $ (\bf{k},\bf{B})$ (or equivalently the vorticity of slow waves is along $\bf{k}\times\bf{B}$, while that of Alfv\'en waves is in the plane $ (\bf{k},\bf{B})$). This question is discussed in the next two subsections, depending on the orientation of the magnetic field. 
In the weak field limit, the efficiencies associated with Alfv\'en and slow waves are independent of the direction of propagation ($Q^{{\rm a}+}\sim Q^{{\rm a}-}$ and $Q^{{\rm s}+}\sim Q^{{\rm s}-}$), because both the Alfv\'en and slow waves are then simply advected and the $(\pm)$ components are undistinguishable.

\subsection{Vertical magnetic field}

A vertical magnetic field does not influence the stationary flow as the fluid flows along the field lines without experiencing any magnetic force. It does however influence the evolution of perturbations that involve some transverse motion, in particular enabling vorticity to propagate. The equations governing the perturbations are described in the Appendix~B. With the magnetic field in the $z$-direction, the directions $x$ and $y$ are equivalent. With no loss of generality we choose the $x$-direction parallel to the transverse wavenumber ($k_{y}=0$) and solve a 2D problem in the plane $ (x,z) $. The boundary conditions at the shock (Eqs~(B13-B14))  impose $\delta v_{y} = \delta B_{y} = 0$. The only waves along this direction would be the Alfv\'en waves, which displacement is along $\bf{k}\times\bf{B}$ (i.e. along $y$ with our choice of axis).  As a consequence, the shock oscillations cannot create Alfv\'en waves, and there are only 4 cycles in this particular field geometry.  Note that the absence of Alfv\'en waves is \emph{not} a consequence of choosing $k_{y}=0$.

In a stationary accretion flow in uniform vertical magnetic field, the ratio of the flow velocity to the Alfv\'en velocity decreases downward, scaling like $v/v_{\rm A}\propto \rho^{-1/2}$. The transition from a superAlfv\'enic flow to a subAlfv\'enic flow is named the Alfv\'en surface. Perturbations cannot be treated in the framework of ideal MHD in the vicinity of this surface because their wavelength becomes infinitely small. Alfv\'en waves can accumulate and be amplified there \citep{williams75}. For the sake of simplicity, we restrict the present study to weak enough magnetic fields so that the flow is superAlfv\'enic everywhere. Given our choice of parameters, the superAlfv\'enic condition requires $v_{{\rm A}z{\rm sh}}/v_{\rm sh} < 0.68$. The study of transAlfv\'enic flows is the subject of a separate study (Guilet et al., in prep.).

\subsection{Horizontal magnetic field}

The equations governing the perturbations in a horizontal magnetic field are described in the Appendix~A. 

In general, both Alfv\'en and slow waves are created at the shock. They propagate mainly along the horizontal magnetic field lines (slow waves also propagate slightly along the perpendicular direction).

The two particular cases $k_x=0$ and $k_y=0$ are simpler because the evolution of perturbations is then planar:

\par (i) if $k_{y}=0$, the boundary conditions at the shock impose $\delta v_{y} = \delta B_{y} = 0$
  while $v_{x}$ and $B_{x}$ are perturbed by the shock oscillations. As $\bf{k}\times\bf{B}$ is the $y$ direction, the shock oscillations do not create Alfv\'en waves but do create slow waves.

\par (ii) if $k_{x}=0$ (${\bf k} \bot {\bf B} $),  the magnetic field lines are not bent and the only magnetic force is the magnetic pressure which adds up to the thermal pressure in the plane $(y,z)$ perpendicular to the field. The evolution of perturbation is thus similar to the hydrodynamical limit, with a modified pressure. Vorticity is advected by the flow together with entropy. This vorticity wave should be viewed as an Alfv\'en wave rather than a slow wave, because it is in the direction of the magnetic field.

\section{The coupling efficiency in the presence of a magnetic field}

In this section we study the effect of the magnetic field on the coupling efficiencies, which are computed by integrating numerically the differential system governing the perturbations, and using the boundary conditions given in the appendices. We consider plane waves with a real frequency which coincide with the real part of the eigenfrequency of a given eigenmode. Growth rates are discussed in Sect. 4.

\subsection{Vertical magnetic field}

\begin{figure}[tbp]
\begin{center}
\includegraphics[width=\columnwidth]{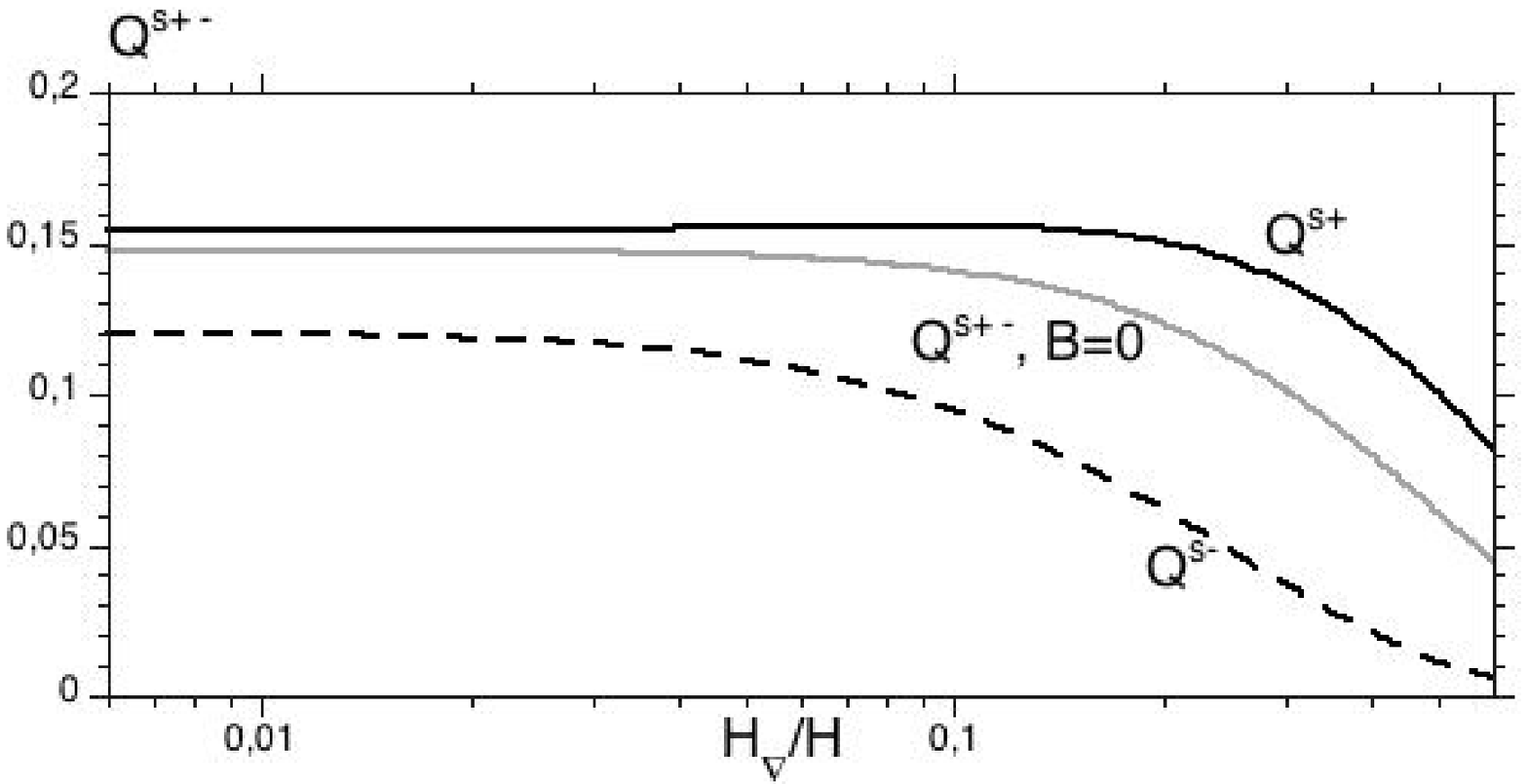}
\includegraphics[width=\columnwidth]{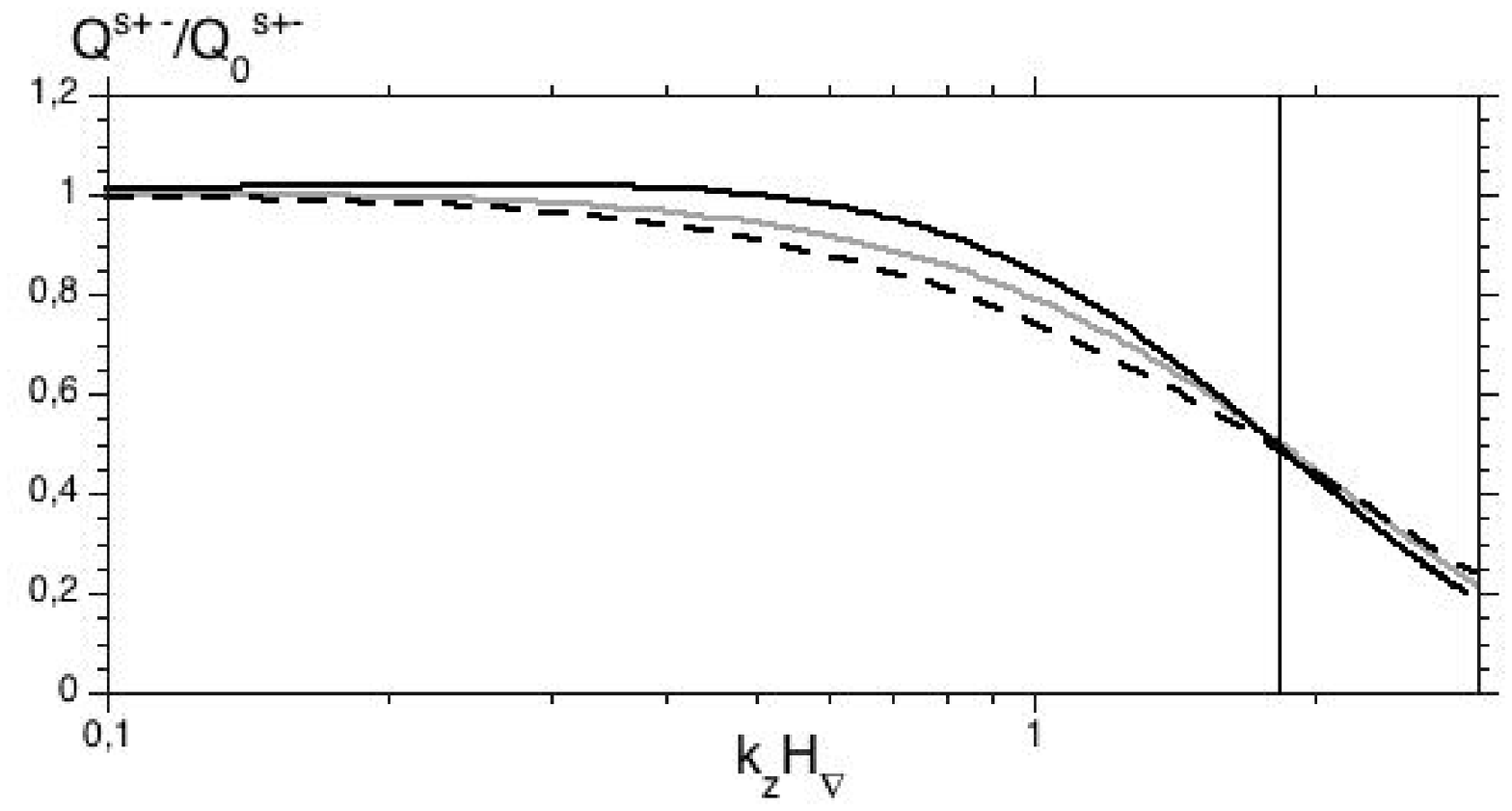}
\caption{\emph{Upper panel:} Efficiency of the slow cycles $Q_{{\rm s}\pm}$ as a function of the height of the potential jump $H_{\nabla}$ in the presence of a vertical magnetic field. The grey line represents the efficiency in the absence of a magnetic field, the black lines correspond to $v_{{\rm A}z{\rm sh}}=0.5v_{\rm sh}$. The dashed line is the (-) wave, the full line the (+) wave. The wave propagating up ($Q_{{\rm s}-}$) is more affected by the cutoff than the wave propagating down ($Q_{{\rm s}+}$). \emph{Bottom panel:} The cycle efficiencies are normalized by $Q_{0}$ defined as the limit of this efficiency when the potential jump is compact. The cutt off happens when $k_{z}H_{\nabla} \gtrsim 1-2$, where $k_z$ is the vertical wavenumber of the slow wave involved in the cycle.}
\label{Qs_dxgrad}
\end{center}
\end{figure}

By solving the evolution of perturbations described in Appendix~A, we remark that the influence of a vertical magnetic field on the individual coupling constants of the different cycles is moderate ($<20\%$) in the long wavelength limit, where the compact approximation is valid. The efficiencies of the two slow cycles at shorter wavelength differ in a way which can be understood in terms of the cutoff described by \cite{foglizzo09}. In the weak field limit, the vertical wavenumber of slow waves can be approximated by:
\begin{equation}
k^{{\rm s}\pm}_{z} \simeq \frac{\omega}{v_{z} \mp v_{{\rm A}z}}.
\label{Bvert_ks}
\end{equation}
The slow wave propagating against the stream (-) has a shorter wavelength than the wave propagating with the stream~(+). It is thus more sensitive to the cutoff induced by the finite scaleheight of the gradients (Fig.~\ref{Qs_dxgrad}, upper panel). As in the hydrodynamic case, the cutoff takes place for $ k_{z}H_{\nabla} \sim 1$ (Fig.~\ref{Qs_dxgrad}, bottom panel).

\subsection{Horizontal magnetic field}

\subsubsection{Straight field lines (${\bf k}\perp{\bf B}$)}

\begin{figure}[tbp]
\begin{center}
\includegraphics[width=\columnwidth]{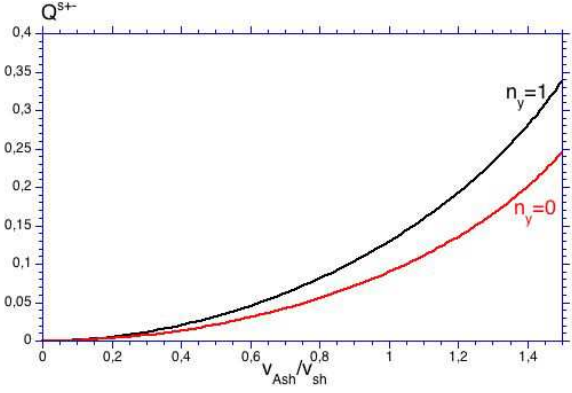}
\caption{Apparition of a slow cycle in the presence of a horizontal magnetic field and ${\bf k}\bot {\bf B}$. This cycle carries no vorticity. It is only a perturbation of the magnetic field and density in such a proportion that the total pressure is not perturbed. }
\label{Qs_apparition}
\end{center}
\end{figure}

We observe the presence of two slow cycles, which were absent in the non-magnetic case (Fig.~\ref{Qs_apparition}). These slow waves taken together (+ and - are equivalent here as these slow waves do not propagate) do not contain any velocity or vorticity perturbation: they are simply a perturbation of the magnetic field strength and density in such a proportion that the total pressure is unperturbed. As their efficiency depends weakly on the transverse wavenumber, we focus on the simplest case of $n_{y}=0$. In this case, the presence of the slow waves can be explained using the conservation of magnetic flux, which perturbation $\delta A \equiv  \delta B_{x}/B - \delta\rho/\rho $ should vanish at the shock (Eqs.~(\ref{Ash})). 

While fast and Alfv\'en waves do not perturb the magnetic flux, the entropy wave is responsible for a perturbation $ \delta A_{S} =  \left(\gamma - 1\right)\delta S/\gamma$. Slow waves have to be created at the shock in order to keep $\delta A_{\rm sh} = 0$.

In the deceleration region, the slow waves necessarily disturb the pressure equilibrium and create an acoustic feedback, in order to satisfy the conservation of both the mass flux and the magnetic flux across the potential jump. This acoustic generation is very similar to the acoustic feedback produced by the deceleration of the entropy wave, required by the conservation of entropy and mass flux across the potential jump.

The effect of the magnetic field on other cycles is only minor and depends on the parameterization of the potential jump. For example the efficiency of the entropy cycle is slightly increased if the compression is kept constant when the magnetic field is varied, while it is slightly decreased if the height of the potential jump is kept constant.

\subsubsection{Bending the field lines: amplification of the vortical cycles \label{amplification}}

\begin{figure}[tbp]
\begin{center}
\includegraphics[width=\columnwidth]{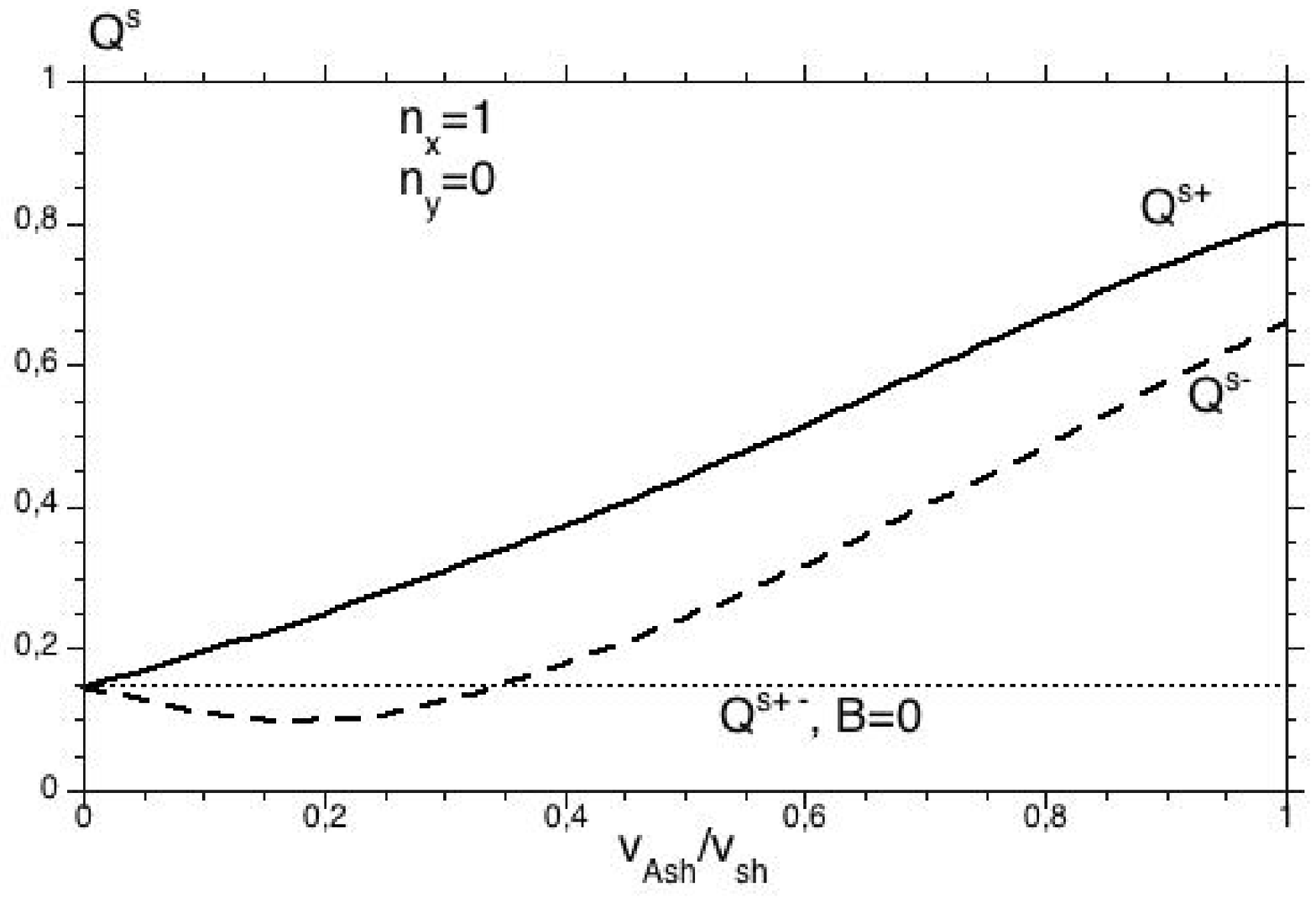}
\includegraphics[width=\columnwidth]{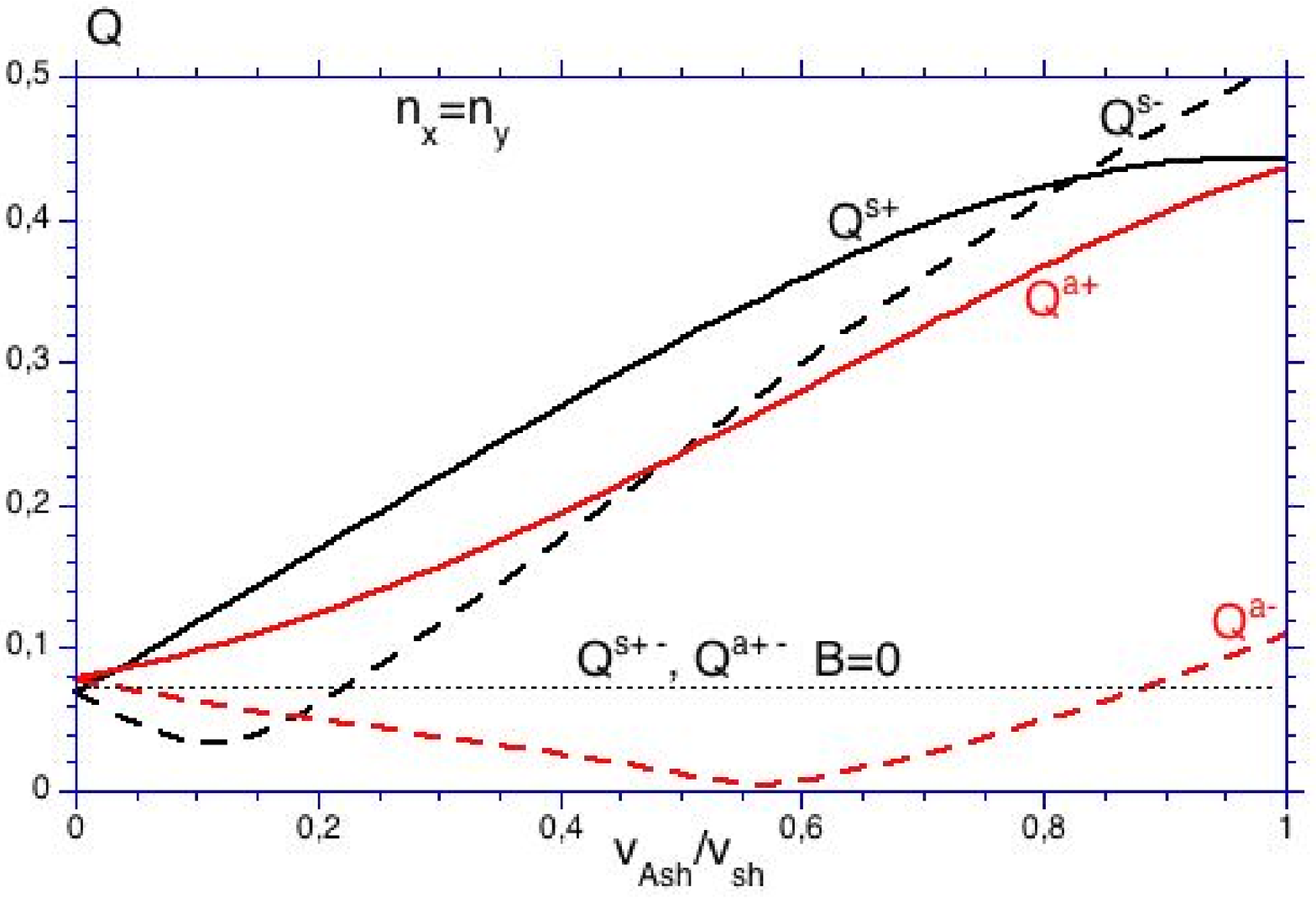}
\caption{Amplification of the vorticity cycles in the presence of a horizontal magnetic field. \emph{Upper panel:} $\bf{k}\parallel\bf{B}$ ($n_{x}=1$, $L_{x}=4$), only slow waves are created at the shock. \emph{Bottom panel:} $\bf{k}$ oblique with respect to $\bf{B}$ ($n_{x}=n_{y}=1$, and $L_{x}=L_{y}=4\sqrt{2}$). Both slow and Alfv\'en waves are created at the shock, and amplified by the presence of the horizontal magnetic field.}
\label{Qs_amplification}
\end{center}
\end{figure}

If $\bf{k}\parallel\bf{B}$, the slow cycles carrying the vorticity are strongly amplified by the presence of a horizontal magnetic field (Fig.~\ref{Qs_amplification}, upper panel). The amplification is almost linear with the magnetic field strength and reaches a factor $\sim 5$ at $v_{{\rm A}x{\rm sh}}=v_{\rm sh}$. Note that this effect is more important than the one described in Sect.~3.2.1, which efficiency only reached $Q^{{\rm s}\pm}\sim 0.1$ at $v_{{\rm A}x{\rm sh}}=v_{\rm sh}$.

When $\bf{k}$ is oblique with respect to $\bf{B}$, the vorticity is distributed in both the Alfv\'en and the slow waves. The Alfv\'en cycles are also amplified, although to a lesser extent than the slow cycles (Fig.~\ref{Qs_amplification}, bottom panel). When varying the angle between $ \bf{k}$ and $\bf{B}$, the total vortical cycle efficiency (the sum of the four slow and Alfv\'en cycles) increases with $k_z/k$.

Although we do not provide an analytical description of this amplification, one can gain insights into its nature by first remarking that this effect is not restricted to slow waves and hence is not due to the compressional nature of the slow waves. Second, this amplification appears only when the field lines are bent by the perturbations. One should however note that a vertical magnetic field bent by perturbations does not lead to the same amplification.

\section{Building of a global mode by adding up cycles}

\subsection{Method}

The cycle efficiencies can be calculated using a complex frequency $\omega=\omega_r+i\omega_i$, where $\omega_r$ is the oscillation frequency and $\omega_i$ the growth rate. Generalizing the analysis of \cite{foglizzo09}, the complex eigenfrequency $\omega$ of a global mode obeys the following relation:
\begin{equation}
Q_{\rm tot} \equiv Q^{\rm S} + Q^{{\rm s}+} + Q^{{\rm s}-} + Q^{{\rm a}+} + Q^{{\rm a}-} + R = 1.\label{defQtot}
\end{equation}
The cycle efficiencies computed in Sect.~3 with a real frequency can be used to characterize marginal stability ($|Q_{\rm tot}|= 1$). $|Q_{\rm tot}| < 1$ indicates a stable mode, and $|Q_{\rm tot}| > 1$ an unstable one.

An estimate of the growth rate associated with a single cycle in the limit $\omega_{i} \ll \omega_{r}$ has been given in \cite{foglizzo09}:
\begin{equation}
\omega_{i} \simeq \frac{1}{\tau_{\rm cycle}}\log{|Q(\omega_{r})|},
\end{equation}
where $\tau_{\rm cycle}$ is the cycle timescale. In the case of an advective-acoustic cycle,
\begin{equation}
\tau_{\rm cycle} =\tau_{\rm aac}\frac{\mu + \M_{\rm sh} }{\mu\left(1+\M_{\rm sh}\right) }
\label{tau_transverse}
\end{equation}
where $\mu$ is defined by $ \mu^{2} \equiv 1 - k_{x}^{2}c^{2}\left(1 - \M^{2}\right)/\omega^{2} $ and $\tau_{\rm aac}$ is the radial advective-acoustic time: 
\begin{equation}
\tau_{\rm aac} \equiv H /(|v_{\rm sh}|\left(1-\M_{\rm sh} \right)).
\end{equation}

The situation is more complicated if several cycles with different timescales are involved in the instability. One may however propose an approximate relation:
\begin{equation}
\omega_{i} \simeq \frac{1}{\tau_{\rm eff}}\log{|Q_{\rm tot}(\omega_{r})|},
\label{wi_Qtot}
\end{equation}
where $\tau_{\rm eff}$ is an effective timescale that would be an average of the different cycle timescales. In the next two subsections we use Eq.~(\ref{tau_transverse}) as a proxy for $\tau_{\rm eff}$ and show that this is a good approximation.

\subsection{Vertical magnetic field \label{Bvert_mode}}

\begin{figure}[tbp]
\begin{center}
\includegraphics[width=\columnwidth]{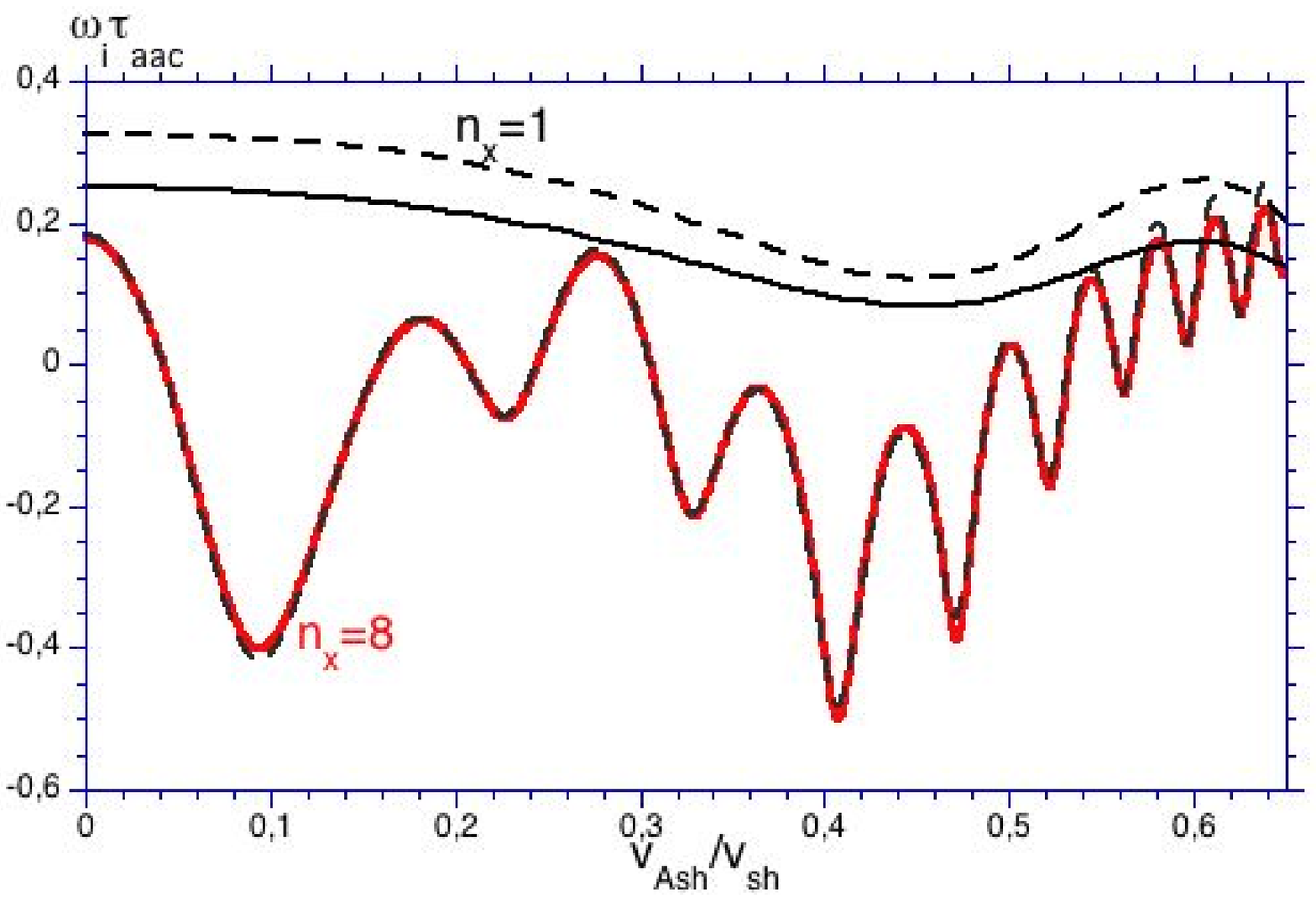}
\includegraphics[width=\columnwidth]{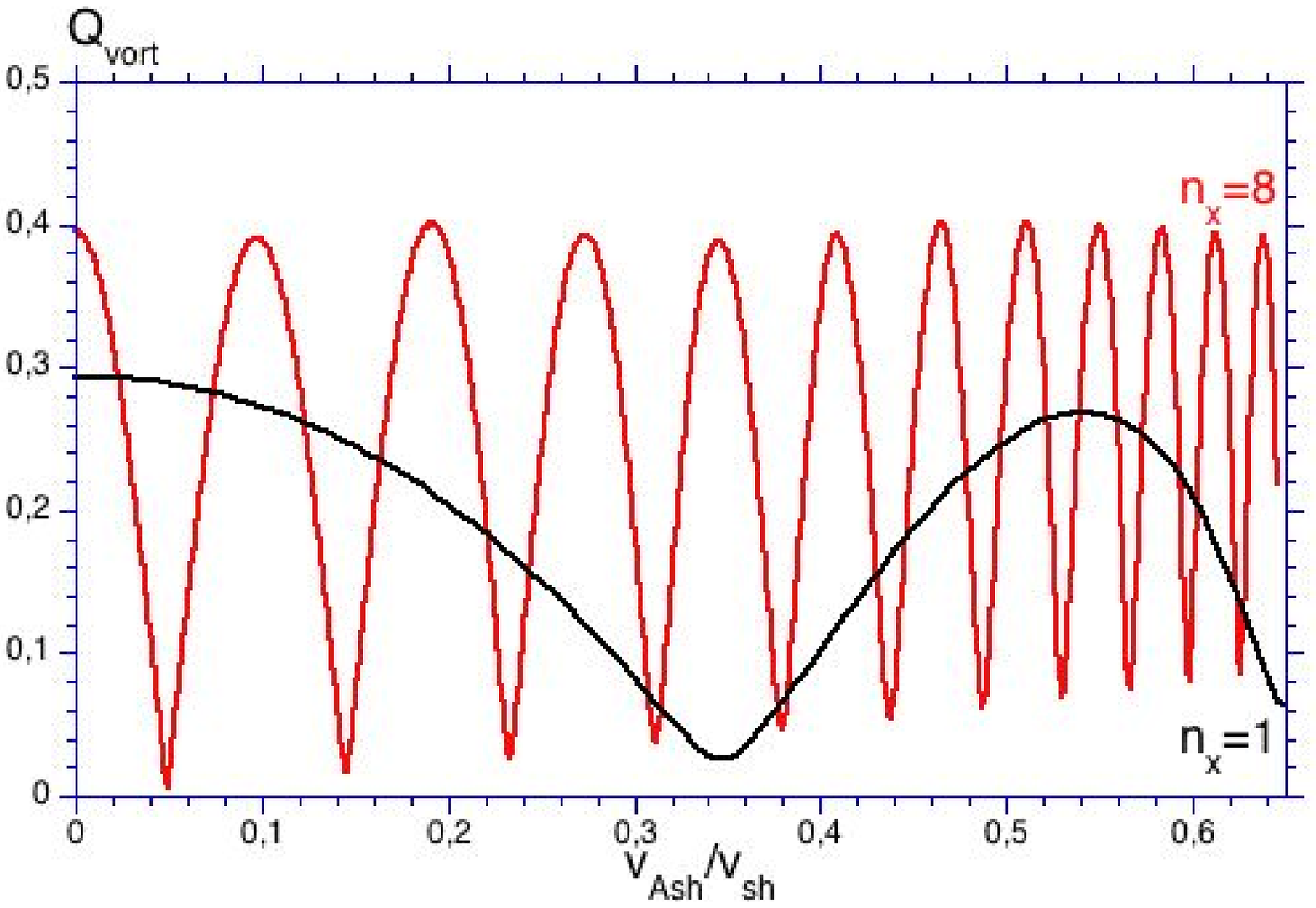}
\caption{Effect of a vertical magnetic field on the modes $n_{x}=1$ (black) and $n_{x}=8$ (grey). \emph{Upper panel:} Growth rate (full line) as a function of the vertical magnetic field strength. The quantity ${\rm log}\left(Q_{\rm tot}\right)/\tau$ is overplotted with dotted lines. \emph{Bottom panel:} Efficiency of the vorticity cycles $Q_{\rm vor}$. The oscillations are due to interferences between the downward propagating (+) and upward propagating (-) slow waves. }
\label{Bvert_wi}
\end{center}
\end{figure}

Fig.~\ref{Bvert_wi} shows the evolution of the most unstable $n_{x}=1$ and $n_{x} = 8$ modes with a vertical magnetic field. The estimate of the growth rate using $Q_{\rm tot}$ (Eq.~(\ref{wi_Qtot})) is in excellent agreement with the eigenvalue if $n_{x}=8$, while the agreement is only reasonable if $n_{x}=1$ (in which case the condition $\omega_{i} \ll \omega_{r}$ is less justified).  This justifies a posteriori our estimate of the effective timescale $\tau_{\rm eff}$, and indicates that it is weakly affected by the magnetic field. Some effect of the magnetic field could have been expected since the vorticity is able to propagate through slow waves, but the (s-) wave is decelerated while the (s+) wave is accelerated, so that both effects on $\tau_{\rm eff}$ cancel each other to first-order. Furthermore, the most efficient cycle in our toy model is the entropic-acoustic cycle and at the magnetic field strength considered, the propagation speed of the fast magnetosonic waves is close to the acoustic one.

The growth rate as a function of the magnetic field strength shows oscillations (Fig.~\ref{Bvert_wi}, upper panel), which are reminiscent of the oscillations in the SASI eigenspectrum observed by \cite{foglizzo07} (Fig.~7). These were interpreted as the consequence of a purely acoustic cycle interacting either constructively or destructively with the advective-acoustic cycle. Similarly the present oscillations can be interpreted as the constructive or destructive effect of the vortical cycles (the two slow cycles).  The phase of their cycle efficiencies $Q^{{\rm s}+}$ and $Q^{{\rm s}-}$ varies with the vertical wavevector of the slow waves as $k_{z}^{{\rm s}\pm}H$ (neglecting a variation of the phase shift during the coupling process). Equation~(\ref{Bvert_ks}) shows that the phases of $Q^{{\rm s}+}$ and $Q^{{\rm s}-}$ vary in opposite directions. This results in oscillations in the value of $Q_{\rm vort}$ as well as $Q_{\rm tot}$ (Fig.~\ref{Bvert_wi}). As $v_{\rm A}$ approaches $v_{z}$, the phase of $Q^{{\rm s}-}$ varies faster and faster, which can recognized in the faster and faster oscillations in Fig.~\ref{Bvert_wi}. The slower phase variation of $Q^{{\rm s}+}$  is responsible for the slower background oscillations in the growth rate of the $n_{x}=8$ mode.

Using a first-order expansion of Eq.~(\ref{Bvert_ks}), an estimate of the first minimum of the growth rate can be deduced from the criterion $\left(k_{z}-k_{z0}\right) H = \pi $:
\begin{equation}
 \frac{v_{{\rm A}z{\rm sh}}}{v_{\rm sh}} \sim \frac{\pi v_{\rm sh}}{\omega H} \sim \frac{1}{2h\left(1-\M_{\rm sh}\right)} ,
\end{equation}
where the second estimate is obtained by approximating the h-th harmonics frequency by $ \omega = 2h\pi/\tau_{\rm aac} $. This gives $v_{{\rm A}z{\rm sh}}/v_{\rm sh}\sim 0.83$ for the mode $n_{x}=1$, and $v_{{\rm A}z{\rm sh}}/v_{\rm sh} \sim 0.10$ for $n_{x}=8$.

\begin{figure}[tbp]
\begin{center}
\includegraphics[width=\columnwidth]{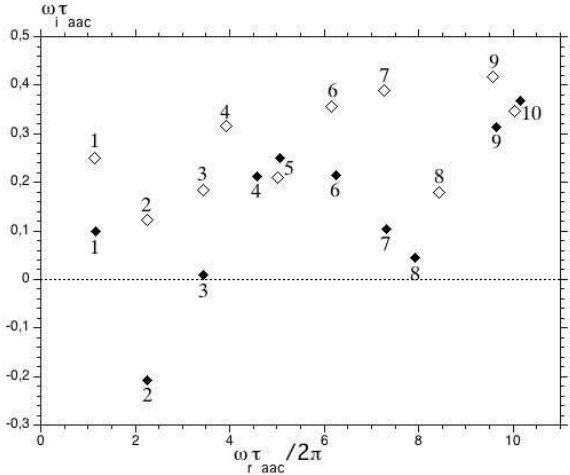}
\caption{Effect of a vertical magnetic field on the eigenspectrum of the compact toy model ($H_\nabla=0$). The most unstable mode is shown for each $n_{x}$ from 1 to 10. White diamonds are without magnetic field, black diamonds with $v_{{\rm A}z{\rm sh}}=0.5v_{\rm sh}$. }
\label{Bvert_eigen}
\end{center}
\end{figure}

Because the entropy and the vorticity cycles are roughly in phase in the non-magnetic limit, the phase variations due to the magnetic field induce an overall decrease of $Q_{\rm tot}$ and $\omega_i$ (Figs.~\ref{Bvert_wi} and \ref{Bvert_eigen}). For any strength of the magnetic field, some modes benefit from a constructive interference of the different cycles and are as unstable as in the absence of a magnetic field (e.g. $n_{x} = 5,10$ in Fig.~\ref{Bvert_eigen}). This phase effect thus never stabilizes completely the {\it compact} toy model ($H_\nabla=0$), but is responsible for an irregular eigenspectrum and lowers the \emph{average} growth rate. If the deceleration region is sufficiently extended vertically, the cutoff effect verified in Sect.~3.1 can stabilize high frequency modes. A strong enough magnetic field may thus be able to completely stabilize the flow if the frequency of the constructive cycles exceeds the frequency cutoff induced by the finite size of the deceleration region.

\subsection{Horizontal magnetic field \label{Bhor_mode}}

\begin{figure}[tbp]
\begin{center}
\includegraphics[width=\columnwidth]{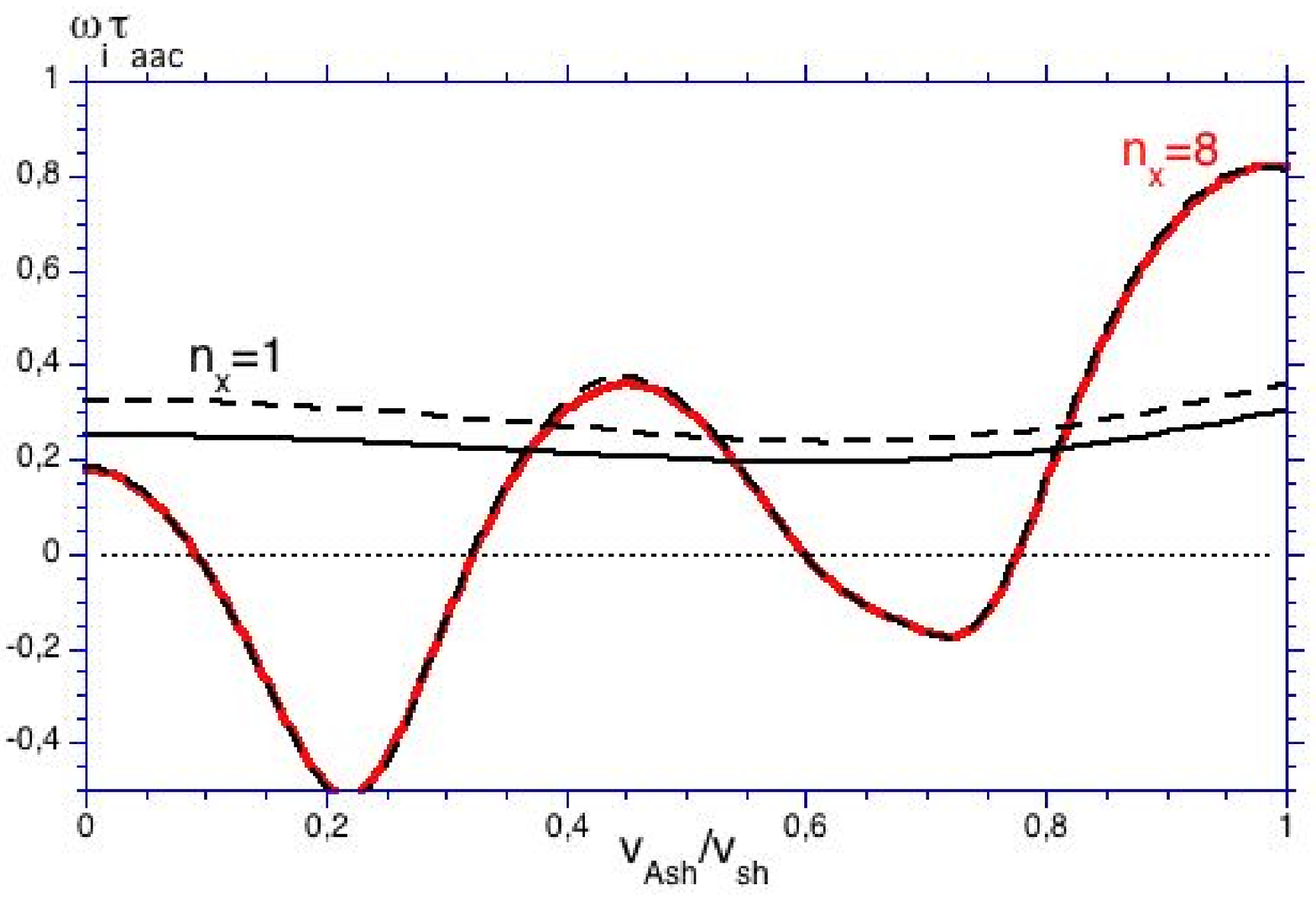}
\includegraphics[width=\columnwidth]{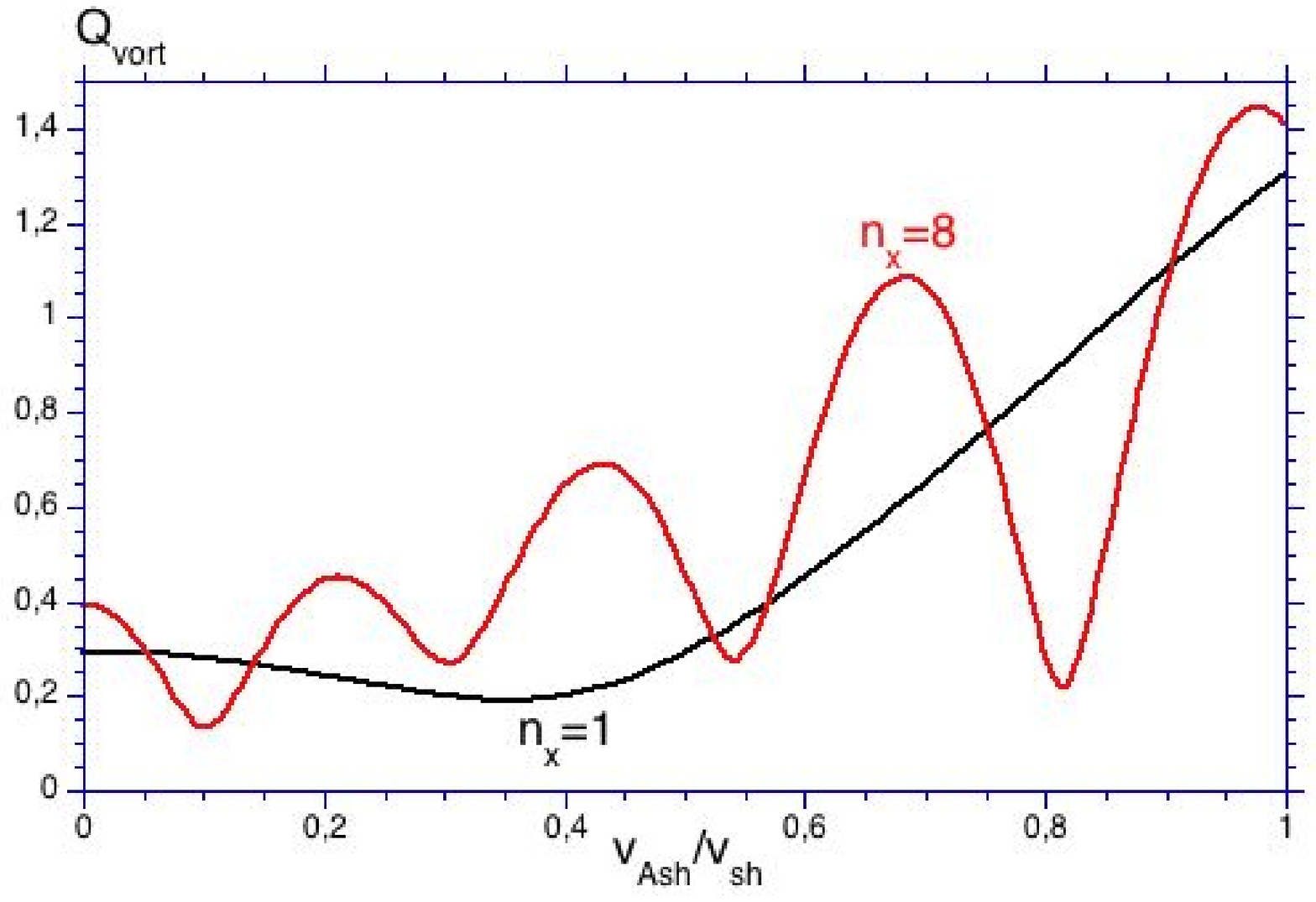}
\caption{Same as Fig.~\ref{Bvert_wi} but with a horizontal magnetic field. The modes presented here have $\bf{k}\parallel\bf{B}$: $n_{x}=1-8$, $n_{y}=0$.}
\label{Bhor_wi}
\end{center}
\end{figure}

Similarly to the case of a vertical magnetic field, the growth rate is well described by Eq.~(\ref{wi_Qtot}) (Fig.~\ref{Bhor_wi}). The oscillations are due to the propagation of vorticity along magnetic field lines, which changes the vertical structure of the slow and Alfv\'en waves as follows:
\begin{equation}
k_{z}^{{\rm s}\pm} \simeq k_{z}^{{\rm a}\pm} =  \frac{\omega}{v}\pm  \frac{k_{x}v_{\rm A}}{v}
\label{Bhor_ks}
\end{equation}
Transverse slow modes are most affected by the magnetic field. The magnetic field strength of the first minimum in the growth rate oscillations, due to the destructive effect of the slow or Alfv\'en cycles on the entropic cycle, can be estimated as: 
\begin{equation}
 \frac{v_{{\rm A}x{\rm sh}}}{v_{\rm sh}} \sim \frac{\pi}{k_{x} H} =  \frac{L_{x}}{2 n_{x} H}
 \label{Bhor_vAmin}
\end{equation}
This gives $v_{{\rm A}x{\rm sh}}/v_{\rm sh} \sim 2$ if $n_{x}=1$, and $v_{{\rm A}x{\rm sh}}/v_{\rm sh} \sim 0.25$ if $n_{x}=8$.

\begin{figure}[tbp]
\begin{center}
\includegraphics[width=\columnwidth]{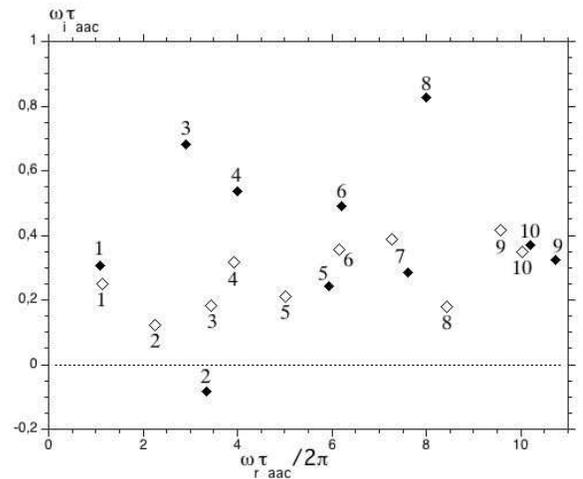}
\caption{Effect of a horizontal magnetic field on the eigenspectrum. The most unstable mode is shown for $n_{x}$ varying from 1 to 10, and $n_{y}=0$. White diamonds are without magnetic field, black diamonds with $v_{{\rm A}x{\rm sh}}=v_{\rm sh}$. }
\label{Bhor_eigen}
\end{center}
\end{figure}

Furthermore, due to the amplification of the vortical cycles described in Sec.~\ref{amplification}, the growth rate at the oscillation maximum increases with the magnetic field strength (Fig.~\ref{Bhor_wi}). As an example, the mode $n_{x}=8$ grows four times faster at $v_{{\rm A}x{\rm sh}} = v_{\rm sh} $ than at $v_{{\rm A}x{\rm sh}} =0$. The eigenspectrum is more irregular than the hydrodynamical one, due to the interferences between the different cycles, and shows higher maximum growth rates: $\omega_{i}\tau_{\rm aac} \sim 0.8$ if $v_{{\rm A}x{\rm sh}}=v_{\rm sh}$, compared to $\omega_{i}\tau_{\rm aac} \sim 0.4$ if $v_{{\rm A}x{\rm sh}}=0$ (Fig.~\ref{Bhor_eigen}).

\section{Discussion and conclusion}

The main results of this study can be summarized as follows:
\begin{enumerate}
\item In the presence of a magnetic field, the advective-acoustic cycle splits in up to 5 different cycles. The acoustic wave becomes a fast magnetosonic wave, while the entropy-vorticity splits into an entropy wave, 2 Alfv\'en waves and 2 slow magnetosonic waves.
\item The acoustic cycle becomes a fast magnetosonic cycle. Its efficiency is not significantly affected by the presence of a magnetic field in any of the configurations considered in this paper.
\item The propagation of the vorticity through slow and Alfv\'en waves leads to a phase difference between the different cycles, which interact either constructively or destructively depending on the mode considered. The consequence of these interferences is a more irregular eigenspectrum.
\item In the superAlfv\'enic regime that we investigate, a vertical magnetic field hardly changes the coupling efficiencies in the compact toy model. It only affects the cutoff associated with the size $H_{\nabla}$ of the deceleration region by changing the vertical structure of the slow waves.
\item The vortical cycles are strongly amplified by a horizontal magnetic field if the field lines are bent ($ \bf{k}.\bf{B} \neq 0 $). Both the Alfv\'en cycles and the slow magnetosonic cycles are amplified. This amplification of the vortical cycles leads to faster growth rates of the modes with $n_x \neq 0$.
\end{enumerate}

These results confirm that the mechanism of the advective-acoustic cycle can be generalized to a magnetized environment, such as core collapse supernovae or possibly the termination shock of pulsar wind nebulae.
In order to estimate the significance of these results for SASI during core collapse, we can estimate the magnetic field strength needed to affect the advective-acoustic cycle. The desynchronization effect studied in Sects.~\ref{Bvert_mode} and \ref{Bhor_mode} requires $v_{\rm Ash} \sim v_{\rm sh}$ for low frequency, low $n_x$ modes. Eq.~(\ref{Bhor_vAmin}) can be extrapolated to a spherical core collapse as follows:
\begin{equation}
 \frac{v_{\rm A}}{v} \sim  \frac{R_{\rm sh}}{2 \sqrt{l(l+1)} H}.
 \label{sphere_vAmin}
\end{equation}
For the dominant $l=1-2$ modes, with $R_{\rm sh} = 110{\rm km}$ and $H=50{\rm km}$, the desynchronisation of the cycles is expected for $v_{\rm A}/v \sim 0.4-0.8$ .

The magnetic field strength at which the amplification of the vortical cycles becomes significant is typically $v_{\rm Ash} \sim (0.5-1)\times v_{\rm sh}$ in our toy model. A more detailed study should determine whether this amplification takes place at the shock or in the region of deceleration. 

Interestingly the magnetic effects described in this paper seem to be significant when the Alfv\'en speed is comparable with the advection speed rather than when the magnetic pressure is comparable with the thermal pressure. These two criteria are related through the Mach number $\M$:
\begin{equation}
\frac{P_{\rm mag}}{P_{\rm th}} = \frac{\gamma}{2}\M^{2}\left( \frac{v_{\rm A}}{v}\right)^{2}.
\end{equation}
In the subsonic flow the pressure ratio is thus smaller than $v_{\rm A}/v$, by a factor $\sim 0.06$ at the shock if $\M_{\rm sh} \sim 0.3$, and as low as $10^{-3}$ if $\M\sim0.05$ at the coupling radius. This raises the possibility that the magnetic field may affect the growth of SASI through the magnetic tension, even if the field is so weak that the magnetic pressure is negligible.

The strength of the magnetic field present in the iron core of a star before the collapse is very uncertain. The best estimate to date is: $B_{\phi}\sim 5.10^9{\rm G}$ and $B_{r}\sim 10^6{\rm G}$ \citep{heger05}, which is too weak to have any effect on SASI: indeed, flux conservation during the collapse ($B/\rho^{2/3} = \rm cste$, e.g. \cite{shibata06}) would lead to $v_{\rm Ash}/v_{\rm sh} \sim 0.001$ at the shock ($r_{\rm sh} \sim150{\rm km}$, $\rho_{\rm sh}\sim 10^9{\rm g.cm^{-3}}$), and $v_{\rm A}/v \sim 0.025$ at the proto-neutron star surface ($r_{\rm PNS} \sim50{\rm km}$, $\rho_{\rm PNS}\sim10^{11}{\rm g.cm^{-3}}$).

A reference Alfven speed near the surface of a strongly magnetized proto-neutron star can be estimated by assuming that the fossil magnetic field is strong enough to give birth to a magnetar ($B\sim 5.10^{15}{\rm G}$) by simple conservation of the magnetic flux :
\begin{eqnarray}
\frac{v_{\rm A}}{v} & \sim &0.93 \times \frac{B_{\rm NS}}{5.10^{15}{\rm G}}\left(\frac{4.10^{14}{\rm g.cm^{-3}}}{\rho_{\rm NS}}\right)^{2/3}\left(\frac{r}{50{\rm km}}\right)^{2} \nonumber \\
& & \left(\frac{\rho}{10^{11}{\rm g.cm^{-3}}}\right)^{7/6}\frac{0.3M_{\odot}.s^{-1}}{\dot M}.
\end{eqnarray}
Such a field could be enough to affect SASI significantly if the amplification of the vortical cycle takes place at the coupling radius, which is slightly above the proto-neutron star surface \citep{scheck08}. Conversely, if the amplification takes place at the shock, even a fossil magnetic field corresponding to magnetar strength would have a negligible effect on SASI ($v_{\rm Ash}/v_{\rm sh} \sim 0.027$). Note however that even stronger magnetic fields are sometimes considered in the literature, with Alfv\'en waves propagating \emph{above} the shock \citep{suzuki08}. 

Given the simplicity of our toy model, we cannot expect it to capture more than the first-order effects of the magnetic field on the advective-acoustic cycle. More accurate estimates should consider a more realistic setup that includes radial convergence and non-adiabatic effects, which are expected to affect the relative importance of the entropy and vorticity cycles. These effects could be estimated in the cylindrical geometry used by \cite{yamasaki08}.

The role of the Alfv\'en surface has not been considered in the present study, because the region of acoustic feedback was assumed to be superAlfv\'enic for the sake of simplicity. If the magnetic field were vertical and strong enough, the magnetic extension of the advective-acoustic cycle to MHD cycles would have to take into account the possible coupling processes taking place at the Alfv\'en surface. The physics of transAlfv\'enic accretion flows will be considered in a forthcoming study (Guilet, Fromang \& Foglizzo, in prep.).

The amplification of the magnetic field observed in the numerical simulations by \cite{endeve08} seems to take place in the nonlinear regime of the MHD-SASI instability, and is thus beyond the scope of the present linear study. The topology of the magnetic field considered by \cite{endeve08} is initially vertical (a split monopole), for which our analysis did not reveal any significant magnetic destabilization. 
Based on the current understanding of the nonlinear saturation process of SASI by parasitic instabilities \citep{guilet09}, we would rather anticipate a larger saturation amplitude of SASI if the field were horizontal near the shock, due to i) the higher growth rate of the cycles involving the bending of field lines by vorticity perturbations, ii) the resistance of magnetic tension to the development of parasitic instabilities such as Rayleigh-Taylor and Kelvin-Helmholtz instabilities. As noted in Guilet et al. (2009) however, the Rayleigh-Taylor instability may still be able to develop as a parasitic instability in the direction perpendicular to the magnetic field, without bending the field lines. Besides, the presence of other magnetically induced parasitic instabilities cannot be ruled out.
Altogether, these qualitative statements are insufficient to explain the growth of the magnetic energy observed by \cite{endeve08}.

\acknowledgments
J.G. is thankful to the Physics Department of the University of Central Florida for its hospitality. This work has been partially funded by the Vortexplosion project ANR-06-JCJC-0119. The authors are grateful to the anonymous referee for his comments that improved the clarity of the paper.

\appendix

\section{Horizontal magnetic field \label{Bhor_appendix}  }

\subsection{Stationary flow}
Denoting by $ \M_{{\rm A}1} = v_{{\rm A}1} /{c_1}  $ the ratio of the Alfv\'en speed and the sound speed ahead of the shock, the postshock Mach number is influenced by a horizontal magnetic field as follows:
\begin{equation}
\frac{\M_{1}^{2}}{\M_{\rm sh}^{2}} = \chi^{2}\left\lbrack 1 + \left(\gamma - 1\right)\left(\frac{\M_{1}^{2}}{2} + \M_{{\rm A}1}^{2}  \right) \right\rbrack - \left(\gamma - 1\right)\M_{{\rm A}1}^{2}\chi^{3} - \left(\gamma - 1\right)\frac{\M_{1}^{2}}{2},\label{machB}
\end{equation}
where $\chi = \rho_{\rm sh}/\rho_1$ is the compression factor, given by:
\begin{equation}
\chi  = \frac{ \left\lbrack{ \left( 1 + \frac{\gamma - 1}{2}\M_{1}^{2} + \frac{\gamma }{2}\M_{{\rm A}1}^{2} \right)^{2} + \left(2 - \gamma \right)\left(\gamma + 1\right)\M_{1}^{2}\M_{{\rm A}1}^{2} }\right\rbrack^{1\over2} - \left( 1 + \frac{\gamma - 1}{2}\M_{1}^{2} + \frac{\gamma }{2}\M_{{\rm A}1}^{2} \right)   }{ \left(2 - \gamma \right)\M_{{\rm A}1}^{2} }.\label{chi}
\end{equation}

\subsection{Definition of the variables}
The 7 variables defined in Eqs.~(\ref{dh}-\ref{dS}) are chosen to describe the perturbations. This choice is guided by the quest for the simplest possible differential system in which the stationary flow gradients do not appear. These variables have the advantage of being conserved through a compact potential jump. $\delta A$ is the perturbation of $B/\rho$, which is conserved in the stationary flow due to magnetic flux conservation. $\delta f$ is the perturbation of the Bernoulli invariant and $\delta S$ is the entropy perturbation. $\delta E_{x}$ and $\delta E_y$ are the perturbation of the transverse electric field along the directions $x$ and $y$, normalized by the stationary electric field $E_y = -vB$.  $\delta K_{1}$  describes the velocity along the $y$ direction and is conserved in the absence of magnetic field \citep{yamasaki08}. 
\begin{eqnarray}
 \delta A & \equiv & \frac{\delta B_{x}}{B} - \frac{\delta\rho}{\rho} ,\\
\delta f & \equiv & v\delta v_{z}  + \frac{2c\delta c}{\gamma -1} + v_{\rm A}^{2}\left( 2\frac{\delta B_{x}}{B} - \frac{ \delta \rho}{\rho} \right), \\
 \delta E_x & \equiv & - \frac{\delta B_{y}}{B}, \\
 \bar{\delta v_{x}} & \equiv & \delta v_{x} - \frac{v_{\rm A}^{2}}{v}\frac{\delta B_{z}}{B}, \\
 \delta E_{y}& \equiv & \frac{\delta B_{x}}{B} +  \frac{\delta v_{z}}{v}, \label{dh} \\
 \delta K_{1} & \equiv & i\omega \delta v_{y} - ik_{y}\delta f ,\\
 \delta S & \equiv & \frac{1}{\gamma - 1}\left( \frac{\delta P}{P} - \gamma \frac{\delta\rho}{\rho} \right). \label{dS}
\end{eqnarray}

The usual physical quantities can be expressed as a function of the new variables through the following set of equations: 
\begin{eqnarray}
\frac{\delta B_{x}}{B} & = & \delta E_{y}- \frac{\delta v_{z}}{v} ,\\
\frac{\delta B_{y} }{B} & = & - \delta E_x , \\
\frac{ \delta B_{z} }{B} & = & - \frac{v}{\omega}\left(  k_{x}\delta E_{y}- k_{y}\delta E_x \right), \\
\delta v_{x} & = & \bar{\delta v_{x}} - \frac{v_{\rm A}^{2}}{\omega}\left( k_{x}\delta E_{y}- k_{y}\delta E_x \right), \\
\delta v_{y} & = & \frac{1}{\omega}\left( k_{y}\delta f - i\delta K_{1} \right), \\
\frac{\delta v_{z}}{v} & = & \frac{1}{1+\M_{A}^{2} - \M^{2}}\left\lbrack -\frac{\delta f}{c^{2}}  + \delta S + \left(1+ \M_{A}^{2} \right)\delta E_{y}+  \left(\M_{A}^{2} - 1 \right) \delta A  \right\rbrack ,\\
\frac{\delta \rho}{\rho} & = & \delta E_{y}- \delta A - \frac{\delta v_{z}}{v}, \\
\frac{\delta c^{2}}{c^{2}} & = & \left( \gamma - 1\right)\left\lbrack \delta S + \delta E_{y}- \delta A - \frac{\delta v_{z}}{v} \right\rbrack ,\\
\delta w_{x} & = & \frac{1}{v_{r}} \left\lbrack \delta K_{1}  + ik_{y}\left(c^{2} \frac{\delta S }{\gamma} + v_{\rm A}^{2}\delta A\right) - v_{\rm A}^{2}ik_{x}\delta E_x  \right\rbrack,
\end{eqnarray}
where $w_x$ is the vorticity along the x direction.

\subsection{Differential system}
The differential system governing the evolution of the perturbations in a horizontal magnetic field is:
\begin{eqnarray}
\frac{\p \delta A}{\p z} & = & \frac{i\omega}{v} \delta A + \frac{ik_{x}}{v}\left\lbrack \bar{\delta v_{x}} - \frac{v_{\rm A}^{2}}{\omega}\left( k_{x}\delta E_{y}- k_{y}\delta E_x \right)  \right\rbrack,\\
\frac{\p \delta f }{\p z} & = & \frac{i\omega}{ v}\left(v^{2}\frac{\delta v_{z}}{v} + c^{2}\frac{\delta S}{\gamma} + v_{\rm A}^{2}\delta A\right) + \frac{ik_{x}v_{\rm A}^{2}}{\omega v}\left\lbrack \omega\bar{\delta v_{x}}  -\left(v^{2} + v_{\rm A}^{2}\right)\left( k_{x}\delta E_{y}- k_{y}\delta E_x \right) \right\rbrack,\\
\frac{\p \delta E_x}{\p z} & = & \frac{i\omega}{v} \delta E_x + \frac{ik_{x}}{\omega v} \left( i\delta K_{1} - k_{y}\delta f  \right) ,\\
\frac{\p \bar{\delta v_{x}}}{\p z} & = & \frac{i\omega}{v} \bar{\delta v_{x}} + \frac{ik_{x}}{v}\left\lbrack \left(c^{2} - v_{\rm A}^{2} \right)\frac{\delta v_{z}}{v} + c^{2}\left( \delta A - \delta E_{y}\right)  +\left( 1-\gamma \right)c^{2}\frac{\delta S}{\gamma} \right\rbrack , \\
\frac{\p\delta E_{y}}{\p z} & = & \frac{i\omega}{v}\left(\delta E_{y}- \frac{\delta v_{z}}{v} \right) + \frac{ik_{y}}{\omega v}\left( i\delta K_{1} - k_{y}\delta f\right),\\
\frac{\p \delta K_{1}}{\p z} & = & \frac{i\omega}{v} \delta K_{1} - \frac{k_{x}v_{\rm A}^{2}}{\omega v}\left\lbrack  -k_{y}\omega \bar{\delta v_{x} } - \left(k_{y}^{2}\left(v^{2} + v_{\rm A}^{2}\right) + \omega^{2}\right)\delta E_x  + k_{x} k_{y}\left(v^{2} + v_{\rm A}^{2}\right) \delta E_{y} \right\rbrack, \\
\frac{\p\delta S}{\p z} & = & \frac{i\omega}{v} \delta S. \\
\end{eqnarray}

\subsection{Wave decomposition in a uniform flow}
\begin{table}[htdp]
\label{Bhor_decomposition}
\caption{Decomposition of the perturbations into waves: horizontal magnetic field} 
\begin{center}
\begin{tabular}{l|c|c|}
\tableline
 & Fast $\pm$ & Slow $\pm$  \\
 \tableline

$\delta A$ &  $ -\frac{k_{x}^{2}c^{2}}{\omega_{\rm c}^{2}} \frac{\delta\rho}{\rho} $ &$-\frac{k_{x}c}{\omega_{\rm c}} \frac{\delta v_{x}}{c} $   \\

$\delta f$  &  $\left\lbrack  k_{z}v  \frac{\omega_{\rm c}^{2} - k_{x}^{2}c^{2}}{\omega_{\rm c}\left(k_{z}^{2}+k_{y}^{2}\right)}  + c^{2} + v_{\rm A}^{2}\left( 1 - 2\frac{k_{x}^{2}c^{2}}{\omega_{\rm c}^{2}} \right)     \right\rbrack\frac{\delta \rho}{\rho}$ &$ \left\lbrack  \frac{k_{z}v}{k_{x}c}  \frac{\omega_{\rm c}^{2} - k_{x}^{2}c^{2}}{k_{z}^{2}+k_{y}^{2}}  + \frac{\omega_{\rm c}}{k_{x}}c + \frac{v_{\rm A}^{2}}{c}\left( \frac{\omega_{\rm c}}{k_{x}} - 2\frac{k_{x}c^{2}}{\omega_{\rm c}} \right)     \right\rbrack \frac{\delta v_{x}}{c} $ \\ 

$\delta E_x $  & $  k_{y}  k_{x} \frac{\omega_{\rm c}^{2} - k_{x}^{2}c^{2}}{\omega_{\rm c}^{2}\left(k_{z}^{2}+k_{y}^{2}\right)} \frac{\delta \rho}{\rho}$ &$ k_{y} \frac{\omega_{\rm c}^{2} - k_{x}^{2}c^{2}}{c\omega_{\rm c}\left(k_{z}^{2}+k_{y}^{2}\right)} \frac{\delta v_{x}}{c} $ \\

$\bar{\delta v_{x}}$  &  $\left( \frac{k_{x}c^{2}}{\omega_{\rm c}} +   \frac{v_{\rm A}^{2}k_{x} k_{z}}{v}\frac{\omega_{\rm c}^{2} - k_{x}^{2}c^{2}}{\omega_{\rm c}^{2}\left(k_{z}^{2}+k_{y}^{2}\right)}    \right)\frac{\delta \rho}{\rho}$ &$ \left( c +   \frac{v_{\rm A}^{2}k_{z}}{cv}\frac{\omega_{\rm c}^{2} - k_{x}^{2}c^{2}}{\omega_{\rm c}\left(k_{z}^{2}+k_{y}^{2}\right)}    \right) \frac{\delta v_{x}}{c}   $   \\

$ \delta  E_{y}$  & $\left( 1 - \frac{k_{x}^{2}c^{2}}{\omega_{\rm c}^{2}} +  \frac{k_{z}}{v}\frac{\omega_{\rm c}^{2} - k_{x}^{2}c^{2}}{\omega_{\rm c}\left(k_{z}^{2}+k_{y}^{2}\right)}    \right)\frac{\delta\rho}{\rho}$ &$  \left( \frac{\omega_{\rm c}}{k_{x}c} - \frac{k_{x}c}{\omega_{\rm c}} +  \frac{k_{z}}{k_{x}vc}\frac{\omega_{\rm c}^{2} - k_{x}^{2}c^{2}}{k_{z}^{2}+k_{y}^{2}}    \right) \frac{\delta v_{x}}{c} $   \\

$\delta K_{1}$  & $ ik_{y}\left\lbrack\frac{\omega_{\rm c}^{2} - k_{x}^{2}c^{2}}{k_{z}^{2}+k_{y}^{2}} -   c^{2} - v_{\rm A}^{2}\left( 1 - 2\frac{k_{x}^{2}c^{2}}{\omega_{\rm c}^{2}} \right)     \right\rbrack\frac{\delta \rho}{\rho}$ &$ ik_{y}\left\lbrack \frac{\omega_{\rm c}}{k_{x}c}\frac{\omega_{\rm c}^{2} - k_{x}^{2}c^{2}}{k_{z}^{2}+k_{y}^{2}} -  \frac{\omega_{\rm c}}{k_{x}}c - \frac{v_{\rm A}^{2}}{c}\left( \frac{\omega_{\rm c}}{k_{x}} - 2\frac{k_{x}c^{2}}{\omega_{\rm c}} \right)     \right\rbrack \frac{\delta v_{x}}{c}$   \\

$\delta S$ &  0 &0 \\
\tableline
$k_{z}$ & $\sim \frac{\omega}{c}\frac{\M \mp \mu }{1-\M^{2}} \pm \frac{v_{\rm A}^{2}}{2c^{2}} \frac{ \left( k_{y}^{2} + k_{z0}^{\pm2} \right) c}{\mu\omega}
$ &  $ \sim\frac{\omega}{v}\left\lbrack 1 \pm \frac{k_{x}v_{\rm A}}{\omega} \left( 1 - \frac{v_{\rm A}^{2}}{2c^{2}}\frac{ k_{y}^{2} + \frac{ \omega^{2}}{v^{2}}}{k_{x}^{2} + k_{y}^{2} + \frac{ \omega^{2}}{v^{2}} } \right) \right\rbrack$  \\
\tableline
\tableline
& Alfv\'en  $\pm$ & Entropy \\
 \tableline
$\delta A$ & 0& $ \left(\gamma - 1\right)\frac{\delta S}{\gamma}$   \\
$\delta f$  &$ ik_{y}v              \frac{\delta w_{x}}{k_{z}^{2} + k_{y}^{2}}$& $\left( v_{\rm A}^{2}\left(\gamma - 1\right) + c^{2} \right)\frac{\delta S}{\gamma} $ \\ 
$\delta E_x $  & $ \pm i\frac{k_{z}}{v_{\rm A}}       \frac{\delta w_{x}}{k_{z}^{2} + k_{y}^{2}} $&0  \\
$\bar{\delta v_{x}}$  &  $ \mp i k_{y}\frac{v_{\rm A}}{v}  \frac{\delta w_{x}}{k_{z}^{2} + k_{y}^{2}}$&0   \\
$ \delta  E_{y}$  &$\frac{ik_{y}}{v}       \frac{\delta w_{x}}{k_{z}^{2} + k_{y}^{2}} $&0   \\
$\delta K_{1}$  & $\left( \omega k_{z} + k_{y}^{2}v\right)   \frac{\delta w_{x}}{k_{z}^{2} + k_{y}^{2}}$&  $-ik_{y} \left( v_{\rm A}^{2}\left(\gamma - 1\right) + c^{2} \right)\frac{\delta S}{\gamma}$   \\
$\delta S$ & 0& $\delta S$ \\
\tableline
$k_{z}$ &  $ \frac{\omega}{v} \pm \frac{k_{x}v_{\rm A}}{v}$  &  $\frac{\omega}{v}$\\
\tableline
\end{tabular}
\end{center}
\label{decomposition_Bhor}
\end{table}

In a uniform flow, the perturbations can be decomposed into 7 different waves, which are the eigenvectors of the differential system: an entropy wave, 2 Alfv\'en waves and 4 magnetosonic waves (2 fast and 2 slow). The perturbation associated with each of these waves is given in Table~\ref{Bhor_decomposition}, where $\omega_{\rm c} \equiv \omega -k_zv$ is the frequency of a wave in the frame comoving with the fluid, $ \mu^{2} \equiv 1 - k_{x}^{2}c^{2}\left(1 - \M^{2}\right)/\omega^{2} $, and $ k_{z0}^{\pm} \equiv \omega/c\times\left(\M\mp\mu\right)/\left(1-\M^2\right) $ is the non-magnetic wavenumber of the sound waves. The vertical wavevector of the magnetosonic waves is obtained by numerically solving the fourth order polynomial dispersion relation:
\begin{equation}
\omega_{\rm c}^{4} - \omega_{\rm c}^{2}k^{2}\left( c^{2} + v_{\rm A}^{2}\right) + k^{2} c^{2}k_{x}^{2}v_{\rm A}^{2} = 0.
\end{equation}
The slow waves are then distinguished from the fast waves as follows: $c^{2} < \omega_{\rm c}^{2}/{k^{2}} < c^{2} + v_{\rm A}^{2}$ for the fast waves, and $0 < \omega_{\rm c}^{2}/{k^{2}} <v_{\rm A}^{2}$ for the slow ones ($v_{\rm A}<c$ in the case considered here).  (+) and (-) waves are distinguished by computing the group velocity $\p \omega/\p k_z$ along the $z$ direction. An approximate expression of $k_z$ that is valid in the weak field limit is also given in Table~\ref{Bhor_decomposition}.

The expressions of Table~\ref{Bhor_decomposition} are the coefficients of a matrix which, when multiplied by the following amplitude vector
\begin{eqnarray}
\left\lbrack \left(\frac{\delta \rho}{\rho}\right)^{{\rm f}\pm}, \left(\frac{\delta v_{x}}{c}\right)^{{\rm s}\pm}, \left(\frac{\delta w_{x}}{k_{z}^{2} + k_{y}^{2}}\right)^{{\rm a}\pm}, \delta S\right\rbrack,
\end{eqnarray}
gives the corresponding perturbations. This matrix is inversed numerically in order to determine the amplitude of each wave as a function of the value of the perturbations.

\subsection{Boundary conditions} 
To obtain the boundary conditions at the shock, we use the conservation of energy, momentum, and mass fluxes as well as the continuity of the magnetic field perpendicular to the shock, and of the electric field parallel to the shock, in the frame of the \emph{perturbed} shock. Assuming that the upstream flow is unperturbed, this gives the following boundary conditions as a function of the shock displacement $ \Delta\zeta $: 
\begin{eqnarray}
\delta A_{\rm sh} & = & 0, \label{Ash}\\
\delta f_{\rm sh} & = & i\omega v_{1}\Delta\zeta\left(1-\frac{v_{2}}{v_{1}}\right),  \\
\delta E_{x \rm sh} & = & 0 , \\
\bar{\delta v_{x}}_{\rm sh} & = &  k_{x}\frac{\delta f_{\rm sh}}{\omega} + ik_{x}\frac{v_{{\rm A}2}^{2}}{v_{2}}\Delta\zeta\left(\frac{v_{2}^{2}}{v_{1}^{2}}-1\right) ,\\
& = & ik_{x}\Delta\zeta\left(1-\frac{v_{2}}{v_{1}}\right)\left\lbrack v_{1} -  \frac{v_{{\rm A}2}^{2}}{v_{2}}\left(1+\frac{v_{2}}{v_{1}}\right)    \right\rbrack, \\
\delta E_{y \rm sh} & = & -\frac{i\omega}{v_{2}}\Delta\zeta\left(1-\frac{v_{2}}{v_{1}}\right) ,\\
\delta K_{1{\rm sh}} & = & 0 ,\\
\frac{ \delta S_{\rm sh}}{\gamma} & = & \frac{i\omega}{c^{2}}\Delta\zeta v_{1}\left(1-\frac{v_{2}}{v_{1}}\right)^{2} -  \frac{\p \Phi}{\p z} \frac{\Delta\zeta}{c^{2}} \left(1-\frac{v_{2}}{v_{1}}\right) .
\end{eqnarray}

Below the potential jump, we use a leaking boundary condition, i.e. no wave propagates upward. For this purpose, at $z = z_{\nabla} - 3H_{\nabla}$ where the flow is homogeneous, we decompose the perturbations into waves as described in the previous section. The only wave that can propagate upward is the fast magnetosonic wave $f-$. The boundary condition requires that its amplitude is zero:
\begin{equation}
\left(\frac{\delta \rho}{\rho}\right)^{{\rm f}-} = 0.
\end{equation}

\subsection{Computation of the cycle efficiencies}
The coupling efficiencies at the shock are computed by decomposing into waves the perturbations just below the shock. For a given value of $\omega$, these perturbations are set by the upper boundary condition described above. The coupling efficiency $Q_{\rm sh}^{i}$ is then computed as the ratio of the amplitude of the wave $i$ ($i={\rm s}\pm,{\rm f}+,{\rm a}\pm,{\rm S}$) to the amplitude of the upward propagating fast magnetosonic wave:
\begin{equation}
Q_{\rm sh}^{i} \equiv  {\delta A^{i}_{\rm sh}\over \delta A^{{\rm f}-}_{\rm sh}}.
\end{equation}

To obtain the coupling efficiency $ Q_{\nabla}^{i} $, we need to determine 
in what proportion an upward propagating fast wave ($f-$) must be added to a wave $i$ at the upper boundary at $z_{\rm sh}$, so that the lower boundary condition is respected. 
For this purpose, we successively set the perturbation at the shock to : (i) a wave $i$ (described in table~\ref{Bhor_decomposition}) of amplitude $\delta A^i_{\rm sh} = 1$, (ii) an upward propagating acoustic wave ($f-$ in table~\ref{Bhor_decomposition}) of amplitude  $\delta A^{f-}_{\rm sh} = 1$. We then integrate the differential system from the shock to $z_{\rm low} = z_{\nabla} - 3H_{\nabla}$, where the lower boundary condition is estimated through the amplitude of the upward propagating wave $f-$ wave: $ x \equiv \left(\frac{\delta \rho}{\rho}\right)^{{\rm f}-}_{z_{\rm low}}$. This resulting quantity is called $x^{i}$ for the wave for the case (i) and $x^{f-}$ for case (ii). The leaking boundary at $z_{\rm low} $ is respected for a combination of the waves $i$ and $f-$ (at the shock) if :


\begin{equation}
\delta A^{{\rm f}-}_{\rm sh}x^{f-} + \delta A^{i}_{\rm sh}x^{i} = 0.
\end{equation}
Thus the coupling efficiency can be computed as :
\begin{equation}
Q_{\nabla}^{i} = { \delta A^{{\rm f}-}_{\rm sh}\over \delta A^{i}_{\rm sh}} = -{x^{i} \over x^{f-}}.
\end{equation}

\section{Vertical magnetic field \label{Bvert_appendix}}
In this appendix we write the equations governing the perturbations in the case of a vertical magnetic field, using the same method as described above in the case of a horizontal magnetic field.

\subsection{Definition of the new variables}
\[
\begin{array}{ccc|c|ccc}
\delta h  & \equiv & \frac{\delta \rho}{\rho} +  \frac{\delta v_{z}}{v}   	&& \frac{\delta \rho}{\rho} &=&   \frac{1}{1- \M^{2}}\left\lbrack \frac{\delta f}{c^{2}}   -\M^{2} \delta h - \delta S  \right\rbrack   \\
 \delta f  & \equiv & v\delta v_{z}  + \frac{2c\delta c}{\gamma -1} 	&& \frac{\delta v_{z}}{v} &=& \frac{1}{1- \M^{2}}\left\lbrack -\frac{\delta f}{c^{2}}   + \delta E_{y}  + \delta S \right\rbrack    \\
 \delta E_{x} & \equiv &  \delta v_{x} -v\frac{ \delta B_{x}}{B}     		&& \frac{\delta B_{z}}{B} &=& \frac{k_{x}}{\omega}\delta E_{x}   \\
 \bar{\delta v_{x}} &  \equiv &  \delta v_{x} - \frac{v_{\rm A}^{2}}{v}\frac{\delta B_{x}}{B}   & &  \delta v_{x} &=& \frac{1}{1-\frac{v_{\rm A}^{2}}{v^{2}}}\left( \bar{\delta v_{x}} -  \frac{v_{\rm A}^{2}}{v^{2}} \delta E_{x} \right) \\
 \delta E_{y} & \equiv &  \delta v_{y} -v\frac{ \delta B_{y}}{B}  		&&  \frac{ \delta B_{x} }{B} &=&    \frac{1}{v\left(1-\frac{v_{\rm A}^{2}}{v^{2}}\right)}\left( \bar{\delta v_{x}} -  \delta E_{x} \right) \\   
 \bar{\delta v_{y}} & \equiv & \delta v_{y} - \frac{v_{\rm A}^{2}}{v}\frac{\delta B_{y}}{B}  &&  \delta v_{y} &=&   \frac{1}{1-\frac{v_{\rm A}^{2}}{v^{2}}}\left( \bar{\delta v_{y}} -  \frac{v_{\rm A}^{2}}{v^{2}} \delta E_{y} \right)   \\
 \delta S & \equiv & \frac{1}{\gamma - 1}\left( \frac{\delta P}{P} - \gamma \frac{\delta\rho}{\rho} \right)  && \frac{\delta B_{y} }{B} &=&  \frac{1}{v\left(1-\frac{v_{\rm A}^{2}}{v^{2}}\right)}\left( \bar{\delta v_{y}} -  \delta E_{y} \right)
\end{array}
\]

\subsection{Differential system}
\begin{eqnarray}
\frac{\p\delta h}{\p z} & = &  \frac{i\omega}{v\left(1-\M^{2} \right)}\left\lbrack  \frac{\delta f}{c^{2}} - \M^{2}\delta h - \delta S \right\rbrack  - \frac{ik_{x}}{v\left( 1-\frac{v_{\rm A}^{2}}{v^{2}} \right)} \left\lbrack \bar{\delta v_{x}} - \frac{v_{\rm A}^{2}}{v^{2}}\delta E_{x}  \right\rbrack, \\
\frac{\p \delta f}{\p z} & = &  \frac{i\omega v}{1-\M^{2}}\left\lbrack  - \frac{ \delta f}{c^{2}} + \delta h + \left(\gamma -1 + \frac{1}{\M^{2}}  \right) \frac{\delta S}{\gamma} \right\rbrack  ,  \\
\frac{\p \delta E_{x}}{\p z} & = &  \frac{i \omega}{v\left( 1-\frac{v_{\rm A}^{2}}{v^{2}} \right)} \left\lbrack \delta E_{x} - \bar{\delta v_{x} }  \right\rbrack,  \\
\frac{\p \bar{\delta v_{x}} }{\p z} & = & \frac{i \omega}{v\left( 1-\frac{v_{\rm A}^{2}}{v^{2}} \right)} \left\lbrack \bar{\delta v_{x}}   -\frac{v_{\rm A}^{2}}{v^{2}}\left( 1 + \frac{k_{x}^{2}\left(v^{2} -v_{\rm A}^{2} \right)}{\omega^{2}} \right) \delta E_{x} \right\rbrack  \nonumber \\
& & +  \frac{i k_{x} v}{1-\M^{2}}\left\lbrack  - \frac{ \delta f}{v^{2}} + \delta h + \left(\gamma -1 + \frac{1}{\M^{2}}  \right) \frac{\delta S}{\gamma} \right\rbrack   , \\
\frac{\p \delta E_{y}}{\p z} & = &  \frac{i \omega}{v\left( 1-\frac{v_{\rm A}^{2}}{v^{2}} \right)} \left\lbrack \delta E_{y} - \bar{\delta v_{y} }  \right\rbrack,  \\
\frac{\p \bar{\delta v_{y}} }{\p z} & = & \frac{i \omega}{v\left( 1-\frac{v_{\rm A}^{2}}{v^{2}} \right)} \left\lbrack \bar{\delta v_{y}}   -\frac{v_{\rm A}^{2}}{v^{2}} \delta E_{y} \right\rbrack ,   \\
\frac{\p\delta S}{\p z} & = & \frac{i\omega}{v} \delta S. \\
\end{eqnarray}

\subsection{Wave decomposition in a uniform flow}
\begin{table}[htbp]
\caption{Decomposition into waves: vertical magnetic field}
\begin{center}
\begin{tabular}{l|c|c|c|c|}
\tableline
 & Fast $\pm$& Slow $\pm$ & Alfv\'en $\pm$ & Entropy \\
 \tableline
$ \delta h $ & $ \left( 1 + \frac{c^{2}}{v^{2}}\frac{k_{z}v}{\omega_{\rm c}} \right) \frac{\delta \rho}{\rho} $ &  $\frac{k_{x}}{\omega_{\rm c}^{2} - k_{z}^{2}c^{2}}\left( \omega_{\rm c} + \frac{c^{2}}{v^{2}}k_{z}v \right)  \delta v_{x} $&0&$\left( 1 -\gamma \right)\frac{\delta S}{\gamma}$ \\
$ \delta f $ & $ c^{2}\left( 1 + \frac{k_{z}v}{\omega_{\rm c}} \right) \frac{\delta \rho}{\rho}$ & $c^{2}\frac{k_{x}}{\omega_{\rm c}^{2} - k_{z}^{2}c^{2}} \left( \omega_{\rm c} + k_{z}v\right)  \delta v_{x}$ &0 &$ c^{2}\frac{\delta S}{\gamma}$ \\
$ \delta E_{x} $ &  $ \frac{\omega_{\rm c}  }{ k_{x}} \left(1- \frac{k_{z}^{2}c^{2}}{\omega_{\rm c}^{2}} \right)\left( 1 + \frac{k_{z}v}{\omega_{\rm c}} \right)    \frac{\delta \rho}{\rho}$ & $\left( 1 + \frac{k_{z}v}{\omega_{\rm c}} \right)  \delta v_{x}$ &0 &0 \\
$ \bar{\delta v_{x}}$ &  $ \frac{\omega_{\rm c}  }{ k_{x}} \left(1- \frac{k_{z}^{2}c^{2}}{\omega_{\rm c}^{2}} \right)\left( 1 + \frac{v_{\rm A}^{2}}{v^{2}}\frac{k_{z}v}{\omega_{\rm c}} \right)  \frac{\delta \rho}{\rho} $ &  $\left( 1 + \frac{v_{\rm A}^{2}}{v^{2}}\frac{k_{z}v}{\omega_{\rm c}} \right)    \delta v_{x}$ &0 &0 \\
$ \delta E_{y}$ & 0&0&$\left(1 \pm \frac{v}{v_{\rm A}} \right) \delta v_{y{\rm a}\pm}$& 0 \\   
$ \bar{\delta v_{y}}$& 0&0 & $\left(1 \pm \frac{v_{\rm A}}{v} \right) \delta v_{y{\rm a}\pm}$& 0  \\
$ \delta S$ & 0 & 0&0&$\delta S$\\
\tableline
$k_{z} $   & $\sim \frac{\omega}{c}\frac{\M \mp \mu }{1-\M^{2}} \pm \frac{v_{\rm A}^{2}}{2c^{2}} \frac{ k_{x}^{2}c}{\mu\omega}
$ & $\sim \frac{\omega}{v \mp v_{\rm A}} \pm \frac{1}{2}\frac{v_{\rm A}^3}{v^3}\frac{\omega^3}{c^2}\frac{v}{k_{x}^2v^2+\omega^2} $ & $ \frac{\omega}{v \mp v_{\rm A}} $ & $\frac{\omega}{v} $\\
\tableline

\end{tabular}
\end{center}
\label{default}
\end{table}%

\subsection{Boundary conditions at the shock}
\begin{eqnarray}
 \delta h_{\rm sh} & = & -\frac{ i \omega \Delta \zeta}{v_{2}}\left( 1 - \frac{v_{2}}{v_{1}}  \right) ,  \\
 \delta f _{\rm sh} & = &  i\omega v_{1} \Delta \zeta \left( 1 - \frac{v_{2}}{v_{1}} \right)  ,\\
 \delta E_{x{\rm sh}} & = &    0   ,   \\
 \bar{\delta v_{x{\rm sh}}} & = & \frac{k_{x}}{\omega}  \delta f _{\rm sh} = ik_{x}v_{1}\Delta \zeta \left( 1 - \frac{v_{2}}{v_{1}} \right)  ,  \\
 \delta E_{y{\rm sh}} & = &   0 ,   \\   
 \bar{\delta v_{y{\rm sh}}} & = & 0 ,\\
 \frac{ \delta S_{\rm sh} }{\gamma} & = & \frac{i \omega v_{1}\Delta\zeta}{c^{2}}\left( 1 - \frac{v_{2}}{v_{1}} \right)^{2} -  \frac{\p \Phi}{\p z} \frac{\Delta\zeta}{c^{2}} \left(1-\frac{v_{2}}{v_{1}}\right) .
\end{eqnarray}

\end{document}